\documentclass[conference, 10pt]{IEEEtran}
\IEEEoverridecommandlockouts
\usepackage{cite}
\usepackage{amsmath,amssymb,amsfonts}
\usepackage{graphicx}
\usepackage{textcomp}
\usepackage{xcolor}
\usepackage{subcaption}
\usepackage{bm}
\usepackage{multirow}
\usepackage{booktabs}
\usepackage{subcaption}
\usepackage{amsthm}\usepackage{balance}\usepackage{url}
\usepackage{algorithm,algpseudocode}
\usepackage{amssymb}

\newtheorem*{proof*}{Proof}

\algrenewcommand\algorithmicrequire{\textbf{Initialization:}}
\algrenewcommand\algorithmicensure{\textbf{Output:}}
\algdef{SE}[DOWHILE]{Do}{doWhile}{\algorithmicdo}[1]{\algorithmicwhile\ #1}%

\makeatletter

\newcommand{\Rmnum}[1]{\expandafter\@slowromancap\romannumeral #1@}
\makeatother

\date{}

\def\BibTeX{{\rm B\kern-.05em{\sc i\kern-.025em b}\kern-.08em
    T\kern-.1667em\lower.7ex\hbox{E}\kern-.125emX}}

\begin{document}
\title{Joint Source-Channel-Check Coding with HARQ for Reliable Semantic Communications
}

\author{Boyuan Li, Shuoyao Wang,~\IEEEmembership{Senior Member,~IEEE}, Suzhi Bi,~\IEEEmembership{Senior Member,~IEEE}, \\
	Liping Qian,~\IEEEmembership{Senior Member,~IEEE}, and Yunlong Cai,~\IEEEmembership{Senior Member,~IEEE} \\
    % Corresponding author.}\\
  \thanks{ 
		  B. Li, S. Wang, and S. Bi  are with the College of Electronic and Information Engineering, Shenzhen University, Shenzhen, China. Liping Qian is with the College of Information Engineering, Zhejiang
		  University of Technology, Hangzhou,  China. 
		  Y. Cai is with the College of Information Science and Electronic Engineering, Zhejiang University, Hangzhou,  China. 
	}
    }
\maketitle
\begin{abstract}\label{abstract}
   % Semantic communication has emerged as a promising paradigm for improving transmission efficiency and task-level reliability.
   % To enhance reliability, recent studies have incorporate  retransmission strategies based on semantic fidelity checking.
 %   However, most existing approaches rely on additional check codewords solely for triggering retransmissions, introducing substantial and often overlooked communication overhead. 
 Semantic communication has emerged as a promising paradigm for improving transmission efficiency and task-level reliability, yet most existing reliability-enhancement approaches rely on retransmission strategies driven by semantic fidelity checking that require additional check codewords solely for retransmission triggering, thereby incurring substantial communication overhead. 
    In this paper,  we propose S3CHARQ, a Joint Source-Channel-Check Coding framework with hybrid automatic repeat request (HARQ) %for efficient and reliable semantic communications.
   % By incorporating the check codeword into the JSCC process, S3CHARQ enables joint optimization of source-channel and check coding (JS3C).
    that fundamentally rethinks the role of check codewords in semantic communications. By integrating the check codeword into the joint source–channel coding (JSCC) process, S3CHARQ enables joint source–channel–check coding (JS3C), allowing the check codeword to simultaneously support semantic fidelity verification and reconstruction enhancement. 
    At the transmitter, %the check codeword is encoded by a semantic fidelity-aware check encoder, making it contains not only fidelity checking capability but also auxiliary information for improved reconstruction quality.
    %At the receiver side, the check codeword is then jointly decoded with the JSCC codeword using a JS3C decoder, as well as independently feed into a LPIPS estimator for perceptual quality estimation.
    a semantic fidelity-aware check encoder embeds auxiliary reconstruction information into the check codeword. At the receiver, the JSCC and check codewords are jointly decoded by a JS3C decoder, while the check codeword is additionally exploited for perceptual quality estimation. 
    %Furthermore, to cope with retransmission misjudgments caused by inherent quality prediction errors, we further propose a reinforcement learning (RL)-based decision module for more accurate sample-level retransmission decisions,
    %enabling adaptive balancing between recovery and refinement information.
    Moreover, because retransmission decisions are necessarily based on imperfect semantic quality estimation in the absence of ground-truth reconstruction, estimation errors are unavoidable and fundamentally limit the effectiveness of rule-based decision schemes. To overcome this limitation, we develop a reinforcement learning (RL)-based retransmission decision module that enables adaptive, sample-level retransmission decisions, effectively balancing recovery and refinement information under dynamic channel conditions.
    Experimental results demonstrate that compared with existing HARQ-based semantic communication systems, 
    the proposed S3CHARQ framework achieves a $2.36$ dB 
    improvement in the 97th percentile PSNR, as well as a $37.45\%$ reduction in outage probability.
\end{abstract}
\section{Introduction}\label{introduction}    

    With the evolution of sixth-generation (6G) communication technologies, immersive real-time multimedia applications, characterized by high throughput requirements, stringent latency constraints, and ultra-high reliability demands, have attracted wide attention.
    Under existing network architectures, the traditional bit-level communication paradigm struggles to simultaneously satisfy these requirements.
    By bridging artificial intelligence (AI) applications with the physical communication layer, semantic communication achieves highly compressed transmission while preserving task-relevant semantic fidelity \cite{Qin2023TOC}.
    Leveraging advances in deep learning, semantic communication (SemComm) systems often use deep neural networks (DNNs) for joint source-channel coding (JSCC) to extract and encode semantic information from various types of sources, such as  text\cite{Qin2021text} and images\cite{2019deepjscc,Yang2025swin,gong2024ICCC}.

    Although SemComm systems exhibit inherent robustness to channel noise,
    systematic designs that ensure end-to-end transmission reliability remain limited, especially for next-generation services such as the metaverse and extended reality.
    Fully guaranteeing reliable semantic information delivery in SemComm systems therefore remains an open challenge. Existing studies have explored this issue from different perspectives, 
    including robust feature extraction networks\cite{Emo2024,topic2025,Knowledge2025}, flexible task-offloading mechanisms\cite{ZhengSem2025}, and large language model (LLM)-based post-processing to recover corrupted semantic features\cite{LLM2025}, all aiming to enhance semantic transmission reliability.
    
    While the aforementioned schemes effectively enhance transmission robustness, feature extraction and encoding in SemComm systems are still primarily conducted at the application and physical (PHY) layers.
    The modern wireless systems rely on HARQ, a cross-layer mechanism spanning both PHY and MAC layers—has proven highly effective in improving transmission reliability, 
    particularly in static or low-mobility scenarios where the channel remains approximately constant over the round-trip time (RTT) of retransmissions.
    Under such conditions, HARQ enables efficient error recovery through retransmission and combining with limited redundancy, 
    making it an attractive reliability-enhancement mechanism for SemComm systems.
    Motivated by these advantages, several recent studies have explored the integration of HARQ into SemComm systems to improve transmission reliability.
    However, existing HARQ-enabled SemComm approaches largely rely on conventional fidelity checking mechanisms  that are designed to guarantee strict bit-wise fidelity.
    These mechanisms are fundamentally mismatched with the inherent tolerance of semantic features to bit errors.
    Therefore, to fully exploit the potential of HARQ in SemComm systems, two fundamental challenges must be addressed.
    \begin{itemize}
        \item \textit{Question 1:  How to accurately assess the impact of channel noise at the semantic-level, rather than at the bit-level?}
        \item \textit{Question 2:  How to jointly optimize retransmission strategies and encoding scheme, improving transmission reliability without sacrificing communication efficiency?}
    \end{itemize}
 
    To address Question 1, several prior works have developed task and modality specific validation schemes. By defining thresholds on perceptual quality metrics, these approaches estimate reconstruction quality at the receiver without requiring access to the original source, 
    and use the estimated quality to guide retransmission decisions.
    For example, \cite{cai2025ccharq} and \cite{leung2023padc} introduced additional reconstruction quality predictors to estimate the Peak Signal-to-Noise Ratio (PSNR)\cite{leung2023padc} and the Structural Similarity Index (SSIM), respectively, and determine whether further retransmissions are required.
    
    To address Question 2, several recent approaches have explored adaptive retransmission strategies with variable codeword lengths to reduce retransmission overhead.
    For instance, \cite{cai2025ccharq} proposed a retransmission scheme CCHARQ, which uses deep reinforcement learning to adaptively adjust the retransmission code length based on SSIM predictions, thereby balancing reconstruction quality and retransmission overhead.
    In contrast, SCBHARQ\cite{liang2025semantic} introduces feature-level negative acknowledgment (NAK) signaling to selectively retransmit different semantic features by comparing the semantic similarity between received noisy features and reference features. 

    Despite these advances, existing studies on encoding and retransmission in semantic communication still exhibit several remarkable limitations:
    i) Due to the heterogeneous optimization goals of error detection and semantic decoding, most current approaches treat these two processes separately, 
    which fails to explore the potential \emph{redundancy between semantic check-coding and semantic source-channel coding}.
    ii) For the retransmission decision-making, existing threshold-based schemes are vulnerable to \emph{prediction errors of reconstruction quality estimators}, severely degrading the reliability of the overall system.
    iii) With respect to the retransmission process itself, retransmissions serve a dual role: on the one hand, 
    they introduce redundancy to mitigate channel noise; on the other hand, they can convey refinement information, functioning similarly to enhancement layers in scalable coding. 
    However, existing retransmission strategies typically rely on binary ACK/NAK signaling, enabling only coarse retransmission decisions. 
    As a result, they lead to static retransmission mechanisms that cannot adapt to the instantaneous distortion level of the received signal, 
    limiting the system's ability to effectively \emph{balance noise recovery and progressive refinement}.
    
    To address these challenges, in this paper, we propose a Joint Source-Channel-Check Coding (JS3C) mechanism, which leverages the redundancy inherent in semantic error detection (SED) to improve image reconstruction performance.
    Based on JS3C, we further  integrate a condition-aware retransmission strategy and optimized retransmission encoding to constitute Source-Channel-Check Coding Hybrid Automatic Repeat reQuest (S3CHARQ). The proposed framework enables more reliable retransmissions and achieves remarkable reconstruction performance under limited retransmission opportunities.
    The main contributions of this paper are summarized as follows:

    \begin{itemize}
        \item {\textcolor{black}{We propose S3CHARQ, a novel HARQ mechanism that jointly optimizes source coding, channel coding, check coding, as well as retransmission coding for SemComm. We propose the JS3C principle, which enables unified optimization of transmission encoding and semantic checking within a single transmission round. 
        		%Building on JS3C, S3CHARQ incorporates a condition-aware retransmission strategy with a recovery-refinement balance, supported by an adaptive retransmission decision module. 
                Overall, 
        		the proposed S3CHARQ framework enables both high-efficiency and high-reliability semantic communications, even over noisy wireless channels.}}
        \item {\emph{Joint Source-Channel-Checking Codecs}}: %To achieve joint optimization of source-channel coding and check coding. At transmitter, a semantic-fidelity-aware (SFA) check encoder generates compact check codeword based on encoded semantic feature, 
        %which is jointly decoded with the received features by the cooperative-aware JSCC (CA-JSCC) decoder to enhance reconstruction quality, as well as independently decoded by LPIPS estimator to estimate the perceptual quality of the current image. 
        %Furthermore, we adopt a multi-stage training strategy based on information bottleneck theory, enabling the check codeword to carry complementary semantic information without compromising its robustness.
        We propose a joint source–channel–check coding (JS3C) design that integrates semantic checking into the JSCC process, enabling unified optimization of source–channel coding and check coding. At the transmitter, a semantic fidelity-aware (SFA) check encoder generates compact check codewords conditioned on encoded semantic features, such that they convey both semantic fidelity verification and auxiliary reconstruction information. At the receiver, the check and JSCC codewords are jointly decoded by a cooperative-aware JSCC (CA-JSCC) decoder to enhance reconstruction quality, while the check codeword is simultaneously exploited by an LPIPS-based estimator for perceptual quality assessment. Furthermore, a multi-stage training strategy grounded in information bottleneck theory is adopted to ensure that the check codeword carries complementary semantic information without compromising robustness. 
        \item {\emph{Condition-aware Retransmission Strategy}}: To cope with prediction errors in receiver-side quality estimation, we design a condition-aware retransmission decision agent based on the Proximal Policy Optimization (PPO) algorithm. By dynamically adjusting the aggressiveness of retransmission based on check codeword, instantaneous channel conditions, and the uncertainty in quality estimation, the proposed agent enables more accurate and robust retransmission decisions under varying distortion levels and channel impairments.
        \item {\emph{Adaptive Recovery-Refinement Re-encoding}}: To mitigate unnecessary redundancy during retransmissions, we incorporate quality estimation feedback into the retransmission encoder and design an entropy optimizer that dynamically adapts redundancy levels for error recovery, while allocating residual bandwidth to refinement information for progressive quality improvement.
        % \item  
    \end{itemize}

    Experiments results validate the superiority of the proposed S3CHARQ. 
         For transmission reliability, S3CHARQ achieves average 37.45\% reduction on outage
        probability, as well as a $2.36$ dB improvement in 97th percentile PSNR, significantly superior compared with existing HARQ-based semantic communication system.  
    In Section~\ref{basic_model_design}, we formulate the problem and describe the proposed system in detail.
    Simulation results are presented in Section~\ref{simulation_results}, and conclusions are drawn in Section~\ref{conclusion}.

\section{Related Works}\label{Related Works}
    \subsection{Semantic Communication}
    With the rapid advancement of deep learning techniques, extensive research has been conducted based on the principle of joint source-channel coding (JSCC) schemes. 
    By leveraging deep neural network (DNN) for semantic feature extraction and representation, JSCC provides an efficient method to address massive data transmission demands while achieving significantly lower reconstruction error\cite{dai2022communication}.
    In recent years, SemComm has been widely studied for text\cite{tang2023text,peng2024robust}, image\cite{erdemir2023generative,gong2025digital}, and video transmission\cite{jiang2022wireless,zhang2023deep}, demonstrating superior communication performance compared with conventional approaches. 
    Furthermore, follow-up studies have continued to improve average transmission performance by incorporating more powerful neural network backbones\cite{Yang2025swin} and introducing variational information bottleneck principles during training\cite{shao2022vib}. 
    As a result, SemComm is increasingly regarded as a promising and effective paradigm for next-generation communication systems.
  
    Despite these remarkable achievements, most existing approaches primarily focus on improving average performance, while relatively few studies explicitly address optimization over the long-tail distribution.
    Training methods based on average-case sampling are inherently limited in their ability to improve the lower performance bound of SemComm systems. 
    Consequently, in practical deployments, these approaches often exhibit pronounced performance fluctuations, failing to reliably satisfy the stringent requirements on worst-case quality and reliability.
    \subsection{Reliable Semantic Communication}
    
    For SemComm systems, achieving reliable transmission is challenging for two reasons. First, optimizing average reconstruction quality alone cannot guarantee the lower bound performance, which is critical for ensuring transmission reliability, especially in worst-case scenarios\cite{choi2025survey}. 
    Second, reliability mechanisms originally designed for bit-level communications, such as check coding and validation, are fundamentally mismatched with the semantic paradigm, 
    which emphasizes semantic-level fidelity rather than strict bit-wise correctness\cite{kalimuthu2025goal}.
    According to classical error control strategies, existing reliability-oriented SemComm approaches can be broadly categorized into two main classes: forward error correction (FEC)-based methods and hybrid automatic repeat request (HARQ)-based methods.
    \subsubsection{FEC-based reliable semantic communication}
    To enable reliable semantic communication, several studies have drawn inspiration from conventional Internet communication protocols (e.g. TCP and RTCP), 
    seeking to adapt and extend bit-level reliability mechanisms to the reliable semantic domain.
    FEC-based methods aim to enhance the robustness of semantic representations by extracting more expressive and noise-resilient features, 
    thereby enabling the receiver to automatically recover errors without requiring retransmissions\cite{hu2022robust,peng2024robust,niu2023deep}. 
    For instance, \cite{hu2022robust} models semantic noise and channel impairments during neural network training, enabling the JSCC encoder to learn robust semantic features, which automatically handling channel noise and semantic distortion at the receiver without additional retransmission rounds.

    Unfortunately, FEC-based schemes typically adopt pre-defined redundancy that is fixed across different channel realizations. 
    As a result, they often fail to guarantee transmission reliability and reconstruction quality\cite{li2025semantic}, 
    especially in time-varying and heterogeneous wireless channel conditions.

    \subsubsection{HARQ-based reliable semantic communication}
    Accordingly, several studies have explored HARQ-based methods to achieve reliable semantic communication.
    To address the mismatch between bit-level error validation and the error-tolerant nature of semantic features, many works have proposed task-oriented validation schemes.
    For instance, \cite{cai2025ccharq} triggers retransmission based on estimated SSIM, and adaptively adjusts retransmission overhead based on the predicted reconstruction quality without requiring access to the original image. 
    \textcolor{black}{To enable more feature-level retransmission decisions, \cite{liang2025semantic} executes validation directly at the semantic-feature level by comparing received features with reference features stored in a semantic codebook, thereby triggering retransmissions based on feature similarity rather than sample-level perceptual quality.} 
    For multi-modal scenario with multiple performance metrics, \cite{wang2025diffusion} jointly considers quality indicators from both text and images, making retransmission decisions based on  a combination of perceptual image quality and semantic text consistency.
    
    Meanwhile, to better exploit corrupted semantic features instead of discarding them, some studies have introduced joint decoding strategies.
    By combining multiple receptions among different transmission rounds, these methods effectively exploit the information contained in noisy features, thereby enhancing data utilization efficiency and overall transmission reliability\cite{cai2025ccharq,TARQ2024,jiang2022sim32}.
    By integrating semantic-level validation with joint decoding into the HARQ framework, these methods can reduce unnecessary retransmissions caused by bit-level errors.

    Despite these advances, most existing HARQ-based SemComm systems rely on predefined thresholds to determine whether retransmission is required. Selecting and adapting such thresholds to current transmission conditions is challenging, as it involves a delicate trade-off between communication efficiency and reconstruction quality. 
    In dynamic wireless environments, where channel conditions and noise levels vary frequently, fixed-threshold schemes struggle to adapt effectively, often resulting in either excessive retransmissions or insufficient error control. 
    Moreover, quality estimates derived from corrupted received signals may deviate significantly from the true reconstruction quality, leading to unnecessary or insufficient retransmissions, further degrading overall system performance.
    
    \subsection{Reinforcement Learning in Semantic Communication}
    Facing the difficulty of accurately quantifying semantic distortion and reconstruction performance, some studies have turned to reinforcement learning (RL) for SemComm. 
    RL-based approaches learn adaptive policies through continuous interaction with the environment, effectively supporting progressive transmission\cite{cai2025ccharq} and cross-layer joint optimization\cite{wang2025cross}, 
    offering a promising approach for addressing the challenges in SemComm systems.
    Moreover, by training agents to learn optimal transmission strategies that adapt to varying channel conditions and noise levels, 
    RL-based methods are particularly well suited for decision-making under dynamic and uncertain wireless environments\cite{sun2025lstm}.

Considering the inherent limitations of static retransmission strategies in HARQ-based semantic communication systems,
recent studies have explored the integration of RL techniques to develop adaptive retransmission mechanisms\cite{huang2024semantic,lu2021reinforcement,liu2025wireless}.
For instance, to jointly optimize energy consumption and data fidelity, \cite{huang2024semantic} formulates semantic communication as a Markov decision process (MDP) and employs a deep Q-network (DQN) agent to determine the optimal number of retransmissions for each sample based on current channel condition, energy consumption, and data reliability.
By designing appropriate reward functions, RL-based methods enable the agent to learn policies that effectively balance the trade-off between communication efficiency and transmission reliability.
However, when retransmission decisions are formulated as a sequential decision-making problem,
system performance becomes highly dependent on how the optimization objectives are encoded in the reward function.
Under diverse channel conditions and heterogeneous reliability requirements,
designing reward functions that accurately reflect retransmission effectiveness and enable robust policy convergence remains a challenging open problem.

\section{System Model}\label{proposed_method}
\subsection{System Model}\label{system_model}
\begin{figure}[!t]
  \centering
  \includegraphics[width=\linewidth,clip,trim={6.8cm 0 8.8cm 0}]{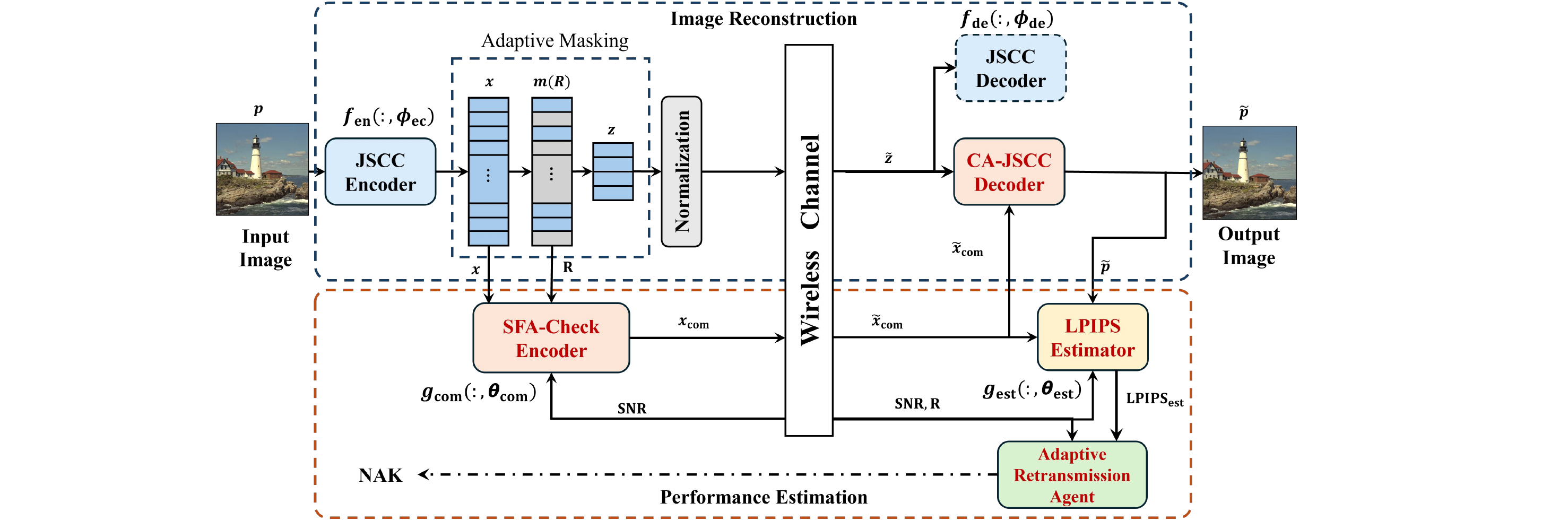}
  \caption{\small{The framework of proposed JS3C.}}
  \label{fig:system_model}
\end{figure}

  As shown in Fig.~\ref{fig:system_model}, we consider a single-user semantic communication (SemComm) system for image transmission. Without loss of generality, the input of the JS3C system is an image $\bm{p} \in \mathbb{R}^{c \times h \times w}$, where $c$, $h$, and $w$ denote the number of channels, height, and width of the image, respectively. The receiver outputs a reconstructed image $\bm{\widetilde{p}} \in \mathbb{R}^{c \times h \times w}$.

In this work, we regard semantic transmissions whose reconstruction quality falls below a certain level as unreliable services, which necessitate retransmission to guarantee task-level reliability. Therefore, accurate quality evaluation of the initially reconstructed samples is essential for triggering retransmissions.
%We adopt the Learned Perceptual Image Patch Similarity (LPIPS) as the perceptual quality metric. 
%However, computing the true LPIPS value requires access to both the original image $\bm{p}$ and the reconstructed image $\bm{\widetilde{p}}$, which is unavailable at the receiver in practical communication scenarios. 
However, in semantic communication systems, the receiver has no access to the original content, rendering exact quality evaluation fundamentally infeasible. 
To address this fundamental limitation, we introduce a performance estimation module that predicts the quality score based on the received check codeword and channel-related information. The estimated score serves as a surrogate quality indicator, enabling retransmission decision-making in the absence of reference images.

  %In this work, we adopt the Learned Perceptual Image Patch Similarity (LPIPS) as the perceptual quality metric. 
 % However, computing the true LPIPS value requires access to both the original image $\bm{p}$ and the reconstructed image $\bm{\widetilde{p}}$, which is infeasible at the receiver side in practical communication scenarios. 
 % To address this challenge, we introduce a performance estimation module that predicts the LPIPS score based on the received check codeword and channel-related information.
 % The estimated $\mathrm{LPIPS}_{\mathrm{est}}$ serves as a surrogate quality indicator at the receiver and enables subsequent decision-making processes under the absence of reference images.
  \subsubsection{Transmitter}
  As shown in Fig.~\ref{fig:system_model}, the transmitter consists of a JSCC encoder, an adaptive masking module, and an \textcolor{black}{SFA check} encoder.
  The semantic encoder extracts high-level features through a learned function $f_{\mathrm{en}} :: \mathbb{R} ^{c \times h \times w}\to\mathbb{R}^{K\times1}$,
  where $K$ denotes the number of encoded symbols.
  Briefly, the semantic encoding function is given by
  \begin{equation}
    \bm{x} = f_{\mathrm{en}}(\bm{p},\bm{\phi}_{\mathrm{en}} ),
    \label{encode_func}
  \end{equation}
  where $\bm{x} \in \mathbb{R}^{K\times1}$ is the encoded semantic features and $\bm{\phi}_{\mathrm{en}}$ denotes the trainable parameters of the semantic encoder.

  To enable the encoded features to adapt to varying communication overhead constraints while preserving robustness to noise under time-varying channel conditions, we introduce an adaptive masking module.
  This module dynamically generates binary mask $\bm{m} \in \mathbb{R}^{K \times 1}$ based on compression ratio $R$ and channel SNR, thereby adjusting the effective length of transmitted semantic features.
  Specifically, for a target compression ratio $R \in \left[0,1\right] $, the number of non-zero elements in the masked feature vector is defined as $K_{\mathrm{mask}} = \left\lceil K \cdot R\right\rceil $, where $\lceil \cdot \rceil$ indicates operation of rounding up. 
  The adaptive masking module then modulates and sorts the encoded features, generating the element $\bm{m_i}$ in mask sequence $\bm{m}$ through the following formulation:
  \begin{equation}
    m_i = \left\{ \begin{aligned}
    &1,   &i\leq K_{\mathrm{mask}}   .\\
    &0,   &\mathrm{else}. 
    \end{aligned}\right.
    \end{equation}
  The masked feature $z \in \mathbb{R} ^{K_{\mathrm{mask}} \times 1}$ is given by
  \begin{equation}
    z = x \odot m,
  \end{equation}
  where $\odot$ indicates the hadmard product. Moreover, in order to generate verification redundancy for performance estimation,
  the proposed \textcolor{black}{SFA check} encoder compresses the encoded features via a learned mapping $g_{\mathrm{com}} :: \mathbb{R}^{K \times 1 } \to \mathbb{R}^{k \times 1}$,
  where $k$ denotes the number of compressed symbols. The compression process is expressed as
  \begin{equation}
    \bm{x}_{\mathrm{com}} = g_{\mathrm{com}}(\bm{x},R,r,\bm{\theta}_{\mathrm{com}}),
    \label{compress}
  \end{equation}
  where $\bm{x}_{\mathrm{com}} \in \mathbb{R}^{k \times 1}$ indicates the encoded verification feature, $r$ denotes the channel state information (CSI) of the physical channel, and $\bm{\theta}_{\mathrm{com}}$ denotes the optimizable parameters.
  For analytical simplicity, the CSI considered throughout this paper is specified by the signal-to-noise ratio (SNR) of the physical channel.

  After power normalization, the normalized signal $\bm{z} \in \mathbb{R}^{K_{\mathrm{mask}} \times 1}$ and $\bm{z}_{\mathrm{com}} \in \mathbb{R}^{k \times 1}$ are transmitted over the physical channel for image reconstruction and performance estimation, respectively.

    \subsubsection{Physical Channel}
    When the encoded signals are transmitted over the physical channel, channel impairments such as noise and multipath fading may distort the transmitted signals.
    Therefore, the received signals could be different from their transmitted signals.
    For example, the received signal $\widetilde{\bm{z}} \in \mathbb{R} ^{K_{\mathrm{mask}} \times 1}$ corresponding to the transmitted signal $\bm{z}$ is given by
    \begin{equation}
      \widetilde{\bm{z}} = \bm{h}\bm{z} + \bm{n},
    \end{equation}
    where $\bm{h}$ denotes the channel gain and $\bm{n}$ denotes the additive channel noise..

    \subsubsection{Receiver}
    After receiving the noisy signals, the receiver reconstructs the source image using the CA-JSCC decoder and estimates the reconstruction quality via an LPIPS estimator.
    For the CA-JSCC decoder, the received verification features $\bm{\widetilde{z}_{\mathrm{com}}}$ and the received masked features $\bm{\widetilde{z}}$ are jointly decoded.
    The corresponding decoding function is defined as $f_{\mathrm{de}} :: \mathbb{R} ^{(K_{\mathrm{mask}} + k)\times 1} \to \mathbb{R}^{c \times h \times w}$,
    and the reconstructed image $\bm{\widetilde{p}} \in \mathbb{R}^{c \times h \times w} $ is obtained as
    \begin{equation}
      \bm{\widetilde{p}} = f_{\mathrm{de}}(\bm{\widetilde{x}}_{\mathrm{com}},\bm{\widetilde{z}},\mathrm{SNR},\bm{\phi}_{\mathrm{de}} ).
      \label{decode}
    \end{equation}

To enable retransmission decisions without access to the original source image, the receiver employs a perceptual quality estimator that provides a surrogate quality score based on the received signals. While conventional metrics such as PSNR and SSIM are widely used for performance evaluation, they are either unbounded or exhibit limited sensitivity in low-quality regimes, making them unsuitable as reliable triggers for retransmission decisions in semantic communication systems.

In this work, we adopt LPIPS as a perceptual quality score for guiding retransmission decisions, as it better reflects semantic fidelity and perceptual degradation under severe channel impairments. Importantly, LPIPS is used solely as a decision signal rather than an optimization target. In Section~V, the effectiveness of the proposed framework is primarily validated using PSNR and subjective visual quality. This further indicates that the performance gains are not achieved by explicitly optimizing for PSNR, but instead arise from a more faithful preservation of semantic information. 

   % To evaluate the image reconstruction quality, the receiver employs an LPIPS estimator, which enables quality assessment without access to the original source image.
    %Notably, due to the unbounded nature of PSNR and the limited sensitivity of SSIM in low-quality regimes, these metrics are inadequate for consistently characterizing perceptual quality from a semantic perspective.
   % To better capture the alignment between semantic features and perceptual reconstruction quality, we adopt LPIPS as the primary evaluation metric and use it to guide subsequent retransmission decisions.
   In particular, by jointly using the verification redundancy $\bm{\widetilde{x}}_{\mathrm{com}}$, received feature $\bm{\widetilde{z}}$, channel condition $r$ and compression ratio $R$,
    the receiver estimates the LPIPS score as
    \begin{equation}
      \bm{\mathrm{LPIPS}}_{\mathrm{est}} = g_{\mathrm{est}}(\bm{\widetilde{z}},\bm{\widetilde{x}}_{\mathrm{com}},r,R,\bm{\theta}_{\mathrm{est}}),
    \end{equation}
    where $\bm{\mathrm{LPIPS}}_{\mathrm{est}}$ denotes the estimated LPIPS value of reconstructed image and the $\bm{\theta}_{\mathrm{est}}$ indicates the trainable parameters of the LPIPS estimator.
\subsection{Retransmission Strategy}
  \label{Retransmission}
\begin{figure}[!t]
    \centering
    \includegraphics[width=\linewidth,clip,trim={2.5cm 0 7.5cm 0}]{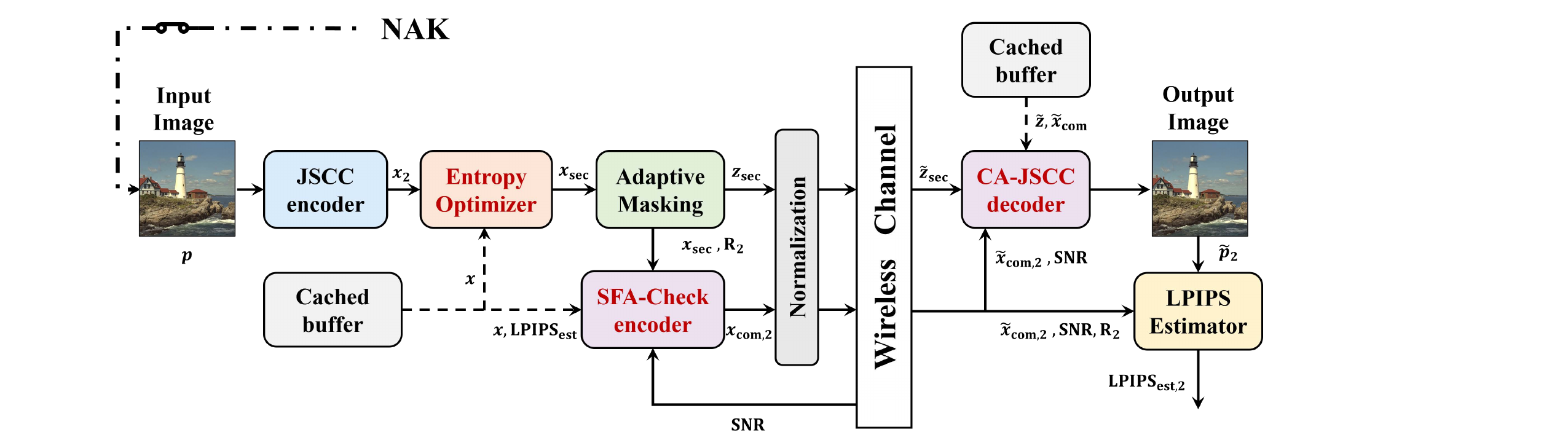}
    \caption{\small{\textcolor{black}{The proposed network architecture of retransmission.}}}
    \label{fig:harq_model}
  \end{figure}

  At the receiver, due to the inherent estimation inaccuracies of neural networks, perfectly quality estimation is generally infeasible.
  Moreover, a key challenge in reliable semantic communication lies in maximizing the coverage of retransmission decisions for severely corrupted images under biased reconstruction,
  which significantly limits the effectiveness of conventional retransmission strategies.
  To address this challenge, we incorporate reinforcement learning and design a PPO-based agent to enable condition-aware retransmission decisions by jointly considering the noise level, compression ratio, and estimated reconstruction quality.
  As shown in Fig.~\ref{fig:system_model}, the adaptive retransmission agent at the receiver collects channel conditions and received verification codewords as state information and makes retransmission decisions at the sample level.
  When retransmission is requested, the agent sends an NAK signal to trigger the retransmission procedure.
  To further meet the reliability requirements of image transmission, we integrate the proposed RL-based retransmission agent into the initial transmission framework shown in Fig.~\ref{fig:system_model},
  together with JS3C, thereby forming the proposed S3CHARQ scheme. This framework supports an adaptive recovery-refinement retransmission process, and the overall system architecture is depicted in the Fig.~\ref{fig:harq_model}.

  As shown in Fig.~\ref{fig:harq_model}, upon receiving a NAK signal from the receiver, the transmitter initiates a retransmission process.
  Specifically, the transmitter performs re-encoding based on the estimated perceptual quality $\bm{\mathrm{LPIPS}}_{\mathrm{est}}$ and the original encoded features $\bm{x}$ to generate retransmission features $\bm{x}_{\mathrm{sec}}$.
  During the retransmission phase, a new mask ratio $R_2$ is applied, and the original features are first re-encoded into $\bm{x}_\mathrm{2}$. 
  Subsequently, an entropy optimizer is employed to further refine the features, yielding $\bm{x}_\mathrm{sec}$. After adaptive masking and SFA-Check encoding, we obtain the corresponding masked feature $\bm{z}_\mathrm{{sec}}$ and corresponding verification redundancy $\bm{x}_{\mathrm{com}, 2}$.
  The entropy optimizing function is expressed as
  \begin{equation}
  \bm{x}_\mathrm{sec} = g_{\mathrm{eo}}(\bm{x}_2, \bm{x}, \bm{\theta}_{\mathrm{eo}}),
  \label{eq:entropy_encoder}
  \end{equation}
  where $\bm{x}_\mathrm{sec}$ denotes the optimized features for retransmission, and $\bm{\theta}_{\mathrm{eo}}$ indicates the trainable parameters of the entropy optimizer.
  For the check-encoding procedure of $\bm{x}_{\mathrm{com}, 2}$, we incorporate the estimated quality $\bm{\mathrm{LPIPS}}_{\mathrm{est}}$ into the encoding procedure, enabling adjustment based on the feedback from the initial reconstruction. The corresponding compressing function is expressed as
  \begin{equation}
    \bm{x}_{\mathrm{com},2} = g_{\mathrm{com},2}(\bm{x},\bm{x}_\mathrm{sec},R_2,r,\mathrm{LPIPS}_\mathrm{est},\bm{\theta}_{\mathrm{com,2}}),
    \label{compress_2}
  \end{equation}
  where $\bm{\theta}_{\mathrm{sec,2}}$ indicates the trainable parameters of the SFA-check encoder in the retransmission phase.
  At the receiver, the CA-JSCC decoder executes joint decoding based on $\bm{\widetilde{x}}_{\mathrm{com}}$, $\bm{\widetilde{z}}$, $\bm{\widetilde{x}}_\mathrm{{{com},2}}$, $\bm{\widetilde{z}}_\mathrm{{sec}}$ and the channel condition $\bm{r}$ to reconstruct the image $\bm{\widetilde{p}}_2$. The joint reconstruction procedure is expressed as
  \begin{equation}
      \bm{\widetilde{p}}_2 = f_{\mathrm{de,2}}(\bm{\widetilde{x}}_{\mathrm{com}},\bm{\widetilde{x}}_{\mathrm{com},2},\bm{\widetilde{z}},\bm{\widetilde{z}}_\mathrm{sec},\mathrm{SNR},\bm{\phi}_{\mathrm{de,2}} ),
      \label{decode}
    \end{equation}
  where $\bm{\phi}_{\mathrm{de,2}}$ denotes the trainable parameters of the CA-JSCC decoder for the retransmission round. 
  For analysis simplicity, we assume that only a single retransmission is permitted and that the channel SNR remains constant during the retransmission process.
  Since the LPIPS prediction of retransmission remains available after retransmission, the proposed framework can be naturally extended to support multiple retransmission rounds,
  thereby accommodating diverse requirements and channel conditions.

\section{Network Architecture}\label{basic_model_design}
In this section, the specific design of proposed system model is explained.
\subsection{System Architecture}
As mentioned above, we adopt the design of SwinJSCC in \cite{Yang2025swin} as the basic JSCC encoder and decoder architecture.
The foundation model is trained end to end using an MSE loss to enable image reconstruction and adaptation to varying channel conditions via the mask ratio.

At the transmitter side, the SwinJSCC encoder $\mathcal{E}_{\mathrm{Swin}}(\cdot)$ encodes the source image $\bm{p}$ into the JSCC-encoded feature $\bm{x}$ as
  \begin{equation}
  \bm{x}
  =
  \mathcal{E}_{\mathrm{Swin}}
  \!\left(
  \mathcal{P}_{p}(\bm{p}),
  \mathcal{P}_{R}(R),
  \mathcal{P}_{\mathrm{snr}}(\mathrm{SNR}),
  \bm{\phi}_{\mathrm{en}}
  \right),
  \label{eq:swin_encoder_arch}
  \end{equation}
  where $\bm{p}$ denotes the input image, $\mathcal{P}_{p}(\cdot)$ represents the patch embedding and feature projection of the image, while $\mathcal{P}_{R}(\cdot)$ and $\mathcal{P}_{\mathrm{snr}}(\cdot)$ denote linear projection and embedding layer that embed the compression ratio and channel SNR into the latent feature space, respectively.
  $\mathcal{E}_{\mathrm{Swin}}(\cdot)$ is implemented by a hierarchical Swin Transformer encoder, and $\bm{\phi}_{\mathrm{en}}$ denotes the trainable parameters of the semantic encoder.
  Following \cite{gong2024imagesem}, the system can be modeled as a Markov chain
   \begin{equation}
    Y \to X \to Z \to \widetilde{Z} \to \widetilde{Y},
    \label{Markov}
   \end{equation}
   where the source image $\bm{p}$ is defined as $Y$, the encoded feature $\bm{x}$ as $X$, and the compressed feature $\bm{x}_{\mathrm{com}}$ as $Z$. After transmission over the wireless channel and decoding, the received signal $\bm{\widetilde{z}}_{\mathrm{com}}$ is $\widetilde{Z}$, and the reconstructed image $\bm{\widetilde{p}}$ is $\widetilde{Y}$.

\subsubsection{SFA Check Encoder}
\begin{figure}[!t]
    \centering
    \includegraphics[width=\linewidth,clip,trim={1cm 3cm 7.6cm 0}]{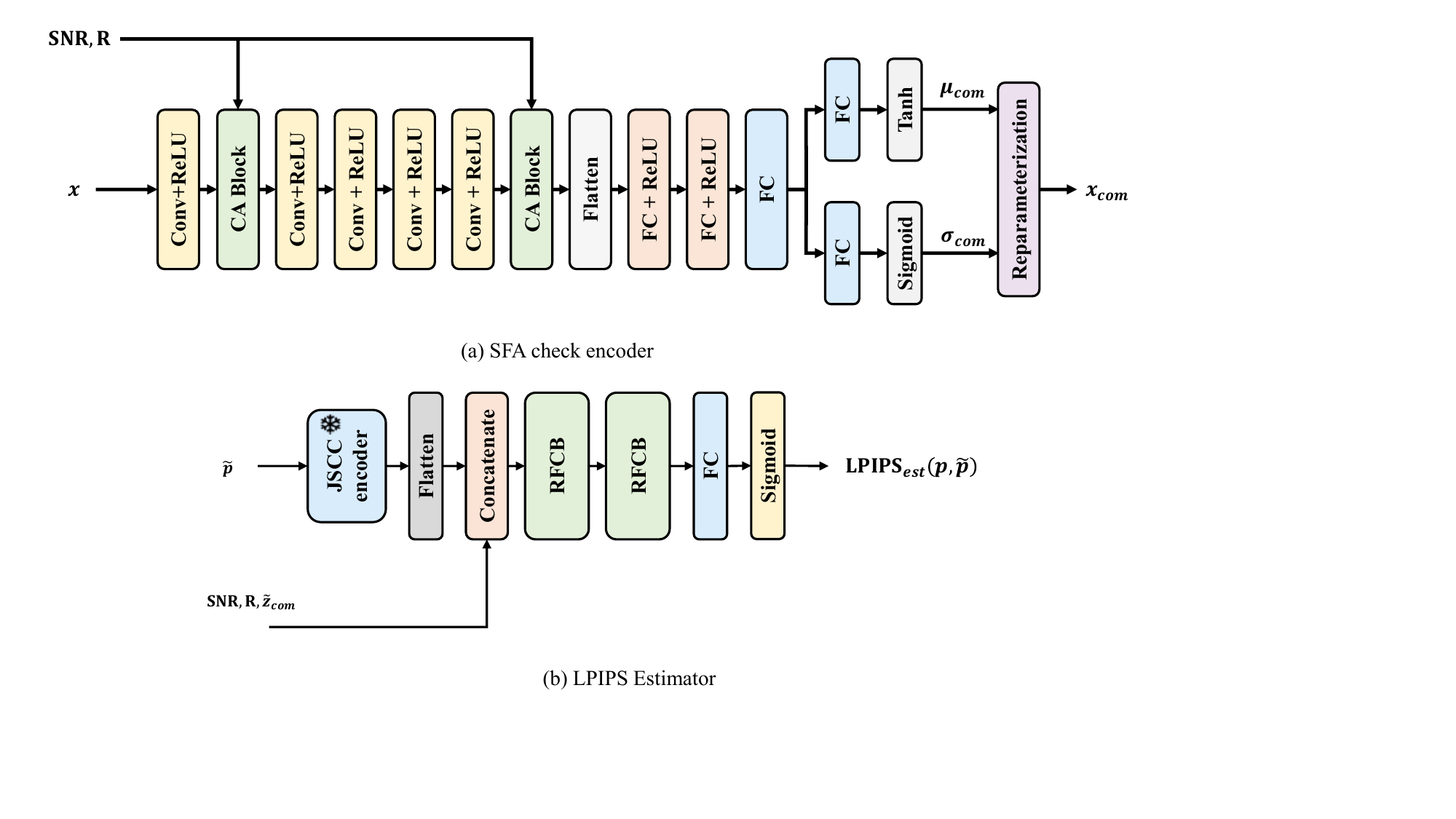}
    \caption{\small{\textcolor{black}{The architecture of SFA check encoder and LPIPS estimator.}}}
    \label{fig:js3c_encoder}
  \end{figure}

  As shown in Fig.~\ref{fig:js3c_encoder}(a), the Semantic-Fidelity-Aware(SFA) check encoder consists of compression and reparameterization blocks.
  Through the SFA check encoder, the encoded feature $\bm{x}$ is further compressed to $\bm{x}_{\mathrm{com}}$, which retains partial semantic information while preserving its inherent error detection capability.
  First, the mean $\bm{\mu}_{\mathrm{com}}$ and variance $\bm{\sigma}_{\mathrm{com}}$ of the compressed feature $\bm{x}_{\mathrm{com}}$ are generated by the SFA check encoder as
  \begin{equation}
  (\bm{\mu}_{\mathrm{com}}, \bm{\sigma}_{\mathrm{com}})
  =
  h_{\mathrm{com}}
  \!\left(
  \bm{x}, R, r, \bm{\theta}_{\mathrm{com}}
  \right),
  \label{eq:check_param}
  \end{equation}
  where $h_{\mathrm{com}}(\cdot)$ denotes the SFA check encoder, parameterized by $\bm{\theta}_{\mathrm{com}}$, which takes the encoded feature $\bm{x}$, the target compression ratio $R$, and the mask ratio $r$ as inputs to generate $\bm{\mu}_{\mathrm{com}}$ and $\bm{\sigma}_{\mathrm{com}}$.
  The compressed feature $\bm{x}_{\mathrm{com}}$ is then obtained through the reparameterization trick as
  \begin{equation}
    \bm{x}_{\mathrm{com}} = \bm{\mu}_{\mathrm{com}} + \bm{\sigma}_{\mathrm{com}} \odot \bm{\epsilon},
  \end{equation}
  where $\bm{\epsilon}$ is sampled from a standard normal distribution $\mathcal{N}(0, \bm{I})$.

  It is crucial for a codec to achieve high decoding performance while minimizing communication overhead. However, these objectives are inherently conflicting: 
  Improving decoding performance typically requires transmitting more symbols, whereas reducing communication overhead degrades noise robustness and thus performance.
  Information Bottleneck(IB) theory \cite{alemi2019ib} provides a principled way to balance this tradeoff. 
  For one-to-one communication, the objective function can be formulated as
   \begin{equation}
    \mathcal{L}_{\mathrm{IB}}(\bm{\phi,\theta}) = -I (Z;Y) + \beta I(Z;X),
   \end{equation}
  where $I(Z;Y)$ indicates the mutual information between the latent feature $Z$ and the decoding output $Y$, which is highly related to decoding performance,
  and $I(Z;X)$ represents the mutual information between the source feature $X$ and $Z$, encouraging reduced communication overhead.
  
  Specifically, the mutual information $I(Z;Y)$ can be written as
   \begin{equation}
    \begin{split}
    I(Z;Y) = \int & p(\bm{y},\bm{z})\log q_{\bm{\theta}}(\bm{y}|\bm{z})\mathrm{d}\bm{y}\mathrm{d}\bm{z} \\
    + & \int p(\bm{y},\bm{z})\log \frac{p(\bm{y}|\bm{z})}{q_{\bm{\theta}}(\bm{y}|\bm{z})}\mathrm{d}\bm{y}\mathrm{d}\bm{z} 
    +  H(Y).
  \end{split}
   \end{equation}

  Following the variational information bottleneck analysis in \cite{gong2024imagesem}, \cite{shao2023extract}, and \cite{peng2019LearningRepresentations}, $I(Z;Y)$ can be optimized by minimizing the MSE loss between the reconstructed image $\bm{\widetilde{p}}$ and the source image $\bm{p}$. The mutual information $I(Z;X)$ can be formulated as
  \begin{equation}
    \begin{split}
    I(Z;X) = \int & p(\bm{z},\bm{x})\log\frac{p_{\phi}(\bm{z}|\bm{x})}{q({\bm{z}})}\mathrm{d}\bm{x}\mathrm{d}\bm{z} \\
    + & \int p(\bm{z},\bm{x})\log\frac{p(\bm{z})}{q({\bm{z}})}\mathrm{d}\bm{x}\mathrm{d}\bm{z}.
  \end{split}
  \end{equation}
  Therefore, $I(Z;X)$ consists of two KL-divergence formulas, $D_{\mathrm{KL}}(p_{\phi}(\bm{z}|\bm{x})||q(\bm{z}))$ and $D_{\mathrm{KL}}(p(\bm{z})||q(\bm{z}))$.
  Since $D_{\mathrm{KL}}(p(\bm{z})||q(\bm{z}))\geqslant0$, the first term dominates and, based on the Markov chain in \eqref{Markov}, can be written as $D_{\mathrm{KL}}(p_{\bm{\theta}_{\mathrm{com}}}(\bm{x}_{\mathrm{com}}\vert \bm{x})\|p(\bm{x}_{\mathrm{com}}))$.
  
  Following \cite{shao2023extract}, the variational marginal distribution $p(\bm{x}_{\mathrm{com}})$ is modeled as a centered isotropic Gaussian distribution $\mathcal{N} (\bm{x}_{\mathrm{com}} | 0,\bm{I})$.
  Therefore, the KL divergence can be simplified to the following form
  \begin{equation}
    \begin{split}
      D_{\mathrm{KL}}&(p_{\bm{\theta}_{\mathrm{com}}}(\bm{x}_{\mathrm{com}}\vert \bm{x})\|p(\bm{x}_{\mathrm{com}}))
      \\= &\sum \left\{\frac{\bm{\mu}_{\mathrm{com}}^2 + \bm{\sigma}_{\mathrm{com}}^2 - 1}{2} - \log \bm{\sigma}_{\mathrm{com}}  \right\}.
    \end{split}
    \label{KL_div}
  \end{equation}

\subsubsection{CA-JSCC Decoder \& LPIPS Estimator}
  At the receiver, to realize efficient joint decoding and quality estimation, we design a Cooperative-Aware JSCC (CA-JSCC) decoder.
  Following \eqref{decode}, the network architecture of CA-JSCC is defined as
  \begin{equation}
    \bm{\widetilde{p}}
    =
    \mathcal{D}_{\mathrm{Swin}}
    \!\left(
    \mathcal{P}_{x}(\bm{\widetilde{x}}_{\mathrm{com}}),
    \mathcal{P}_{z}(\bm{\widetilde{z}}),
    \mathcal{P}_{\mathrm{snr}}(\mathrm{SNR}),
    \bm{\phi}_{\mathrm{de}}
    \right),
    \label{eq:swin_decoder_arch}
  \end{equation}
  where $\mathcal{P}_{x}(\cdot)$, $\mathcal{P}_{z}(\cdot)$, and $\mathcal{P}_{\mathrm{snr}}(\cdot)$ are learnable linear layers that map heterogeneous inputs into a unified latent space for feature alignment.
  $\mathcal{D}_{\mathrm{Swin}}(\cdot)$ denotes the decoder composed of stacked Swin Transformer blocks\cite{Yang2025swin}, parameterized by $\bm{\phi}_{\mathrm{de}}$, 
  which jointly exploits spatial correlations and cross-feature dependencies to reconstruct the target image $\bm{\widetilde{p}}$.
  
  For LPIPS estimation, we define an LPIPS estimator as shown in Fig.~\ref{fig:js3c_encoder}(b), which is trained by minimizing the MSE loss between the ground-truth LPIPS and its estimated value. 
  Considering the coupled nature of LPIPS estimation, SFA check coding, and CA-JSCC decoding. 
  we jointly optimize the decoding procedure and performance estimation stages.
  Specifically, the previously trained SwinJSCC encoder is frozen, while the SFA check encoder, CA-JSCC decoder, and LPIPS estimator are trained jointly using an Information Bottleneck-based loss function defined as
  \begin{equation}
    \mathcal{L} = \frac{1 }{B} \sum_{i = 1}^{B}||(\mathrm{LPIPS}(\bm{p},\widetilde{\bm{p}}),\mathrm{LPIPS_{est}})||_2^2 + \mathcal{L}_{\mathrm{IB}},
  \end{equation}
  where $\mathcal{L}_{\mathrm{IB}}$ is given by
  \begin{equation}
      \mathcal{L}_{\mathrm{IB}} = \frac{1 }{B} \sum_{i = 1}^{B}d(\bm{p},\widetilde{\bm{p}})  +  \gamma D_{KL}(p_{\bm{\theta}_{\mathrm{com}}}(\bm{x}_{\mathrm{com}}\vert \bm{x})\|p(\bm{x}_{\mathrm{com}})). 
    \label{IBloss}
  \end{equation}
  Here, $\gamma$ is the weight of the KL divergence term in the loss function.

\subsubsection{Entropy Optimizer}
\begin{figure}[!t]
    \centering
    \includegraphics[width=\linewidth,clip,trim={0.1cm 0.1cm 0cm 0}]{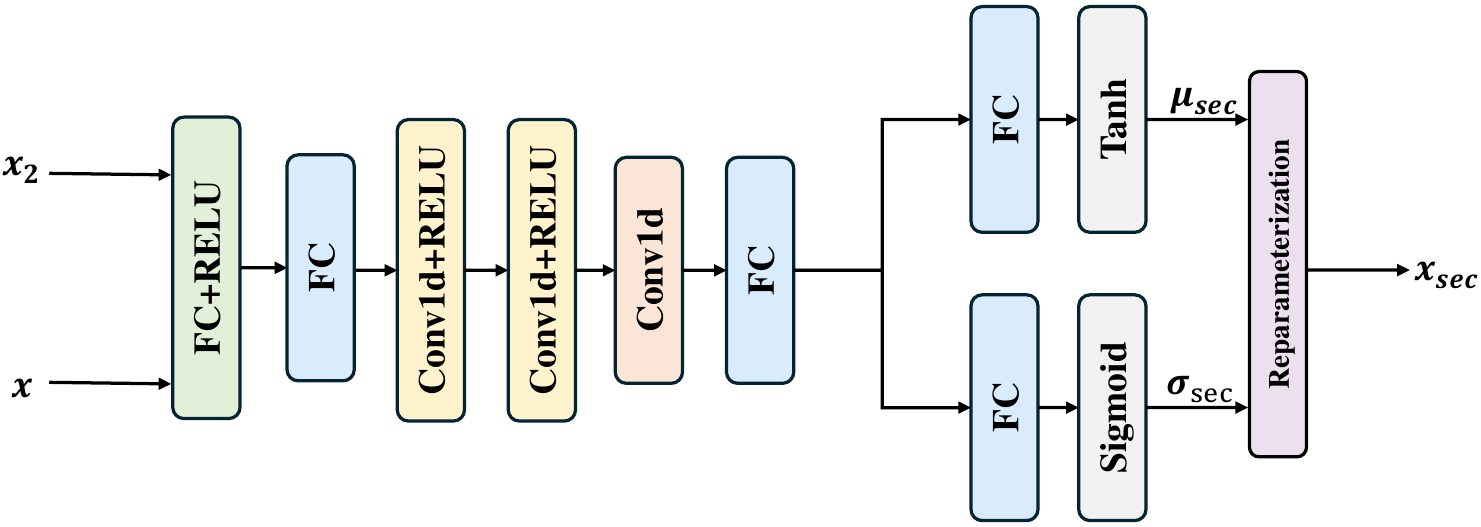}
    \caption{\small{The architecture of the entropy optimizer.}}
    \label{fig:entropy}
  \end{figure}

  As shown in Fig.~\ref{fig:harq_model}, 
  following the design philosophy of the SFA check encoder, we adapt the basic system to the retransmission scenario by reducing the homogeneity between the encoded features used in the initial transmission and the retransmission.
  To this end, we design an Entropy Optimizer, whose basic structure is illustrated in Fig.~\ref{fig:entropy}.

  In addition, to enable the SFA check encoder, CA-JSCC decoder and LPIPS estimator to fully exploit effective information for encoding, decoding, and quality estimation,
  we retrain the corresponding modules. Since the retransmission modules operate independently of the first-round transmission,
  the basic system is frozen, and a loss function similar to \eqref{IBloss} is used to jointly train these modules.
 \begin{figure}[!t]
  	\centering
  	\includegraphics[width=\linewidth,clip,trim={3.5cm 3cm 2.5cm 3cm}]{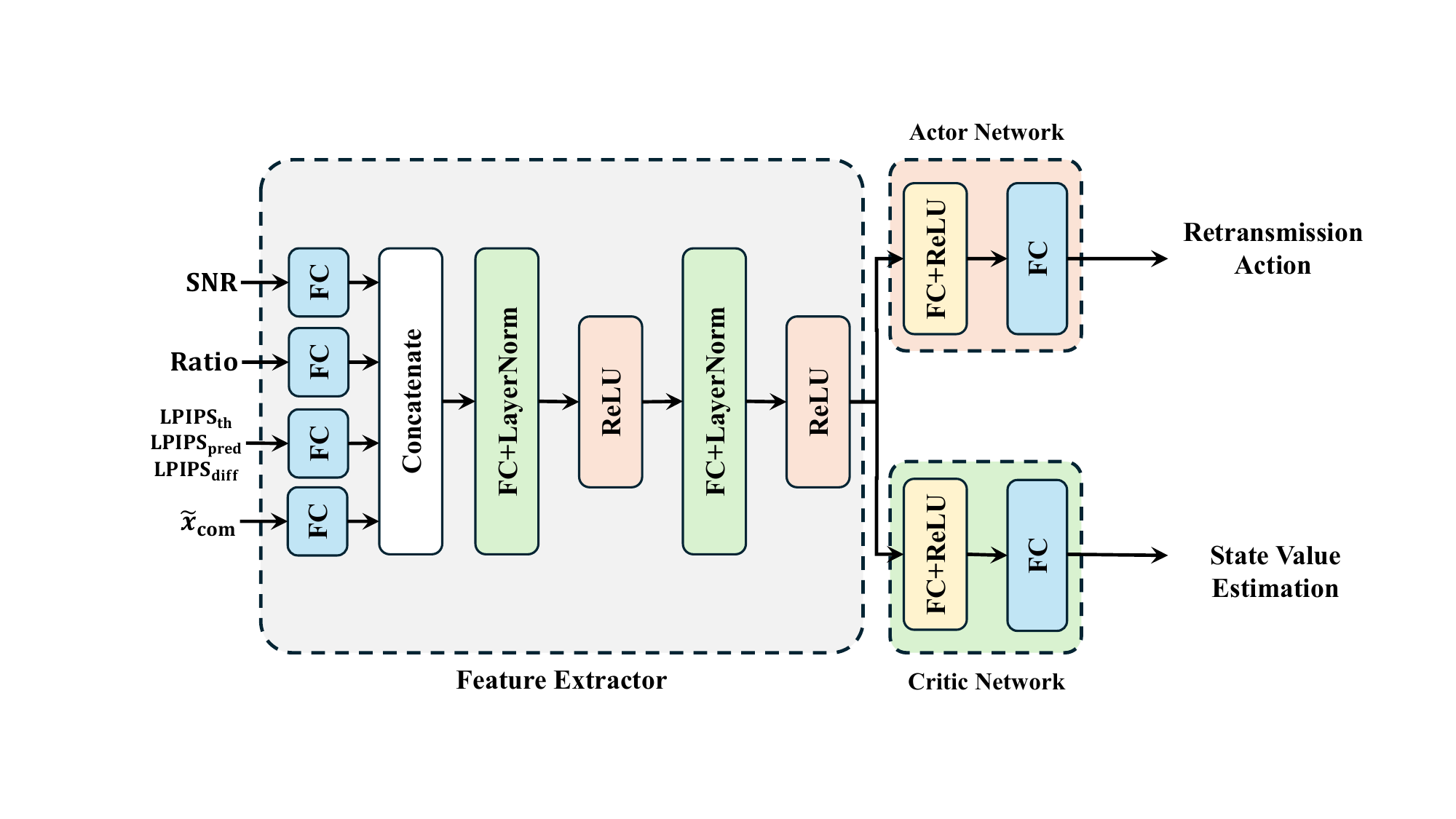}
  	\caption{\small{The architecture of the Adaptive Retransmission Agent.}}
  	\label{fig:RLAgent}
  \end{figure}
\subsection{Adaptive Retransmission Agent}
    As discussed in Section \ref{Retransmission}, we adopt a PPO-based reinforcement learning agent to jointly optimize the actor and critic networks, which is shown in Fig.~\ref{fig:RLAgent}.
    The goal of PPO agent is to learn a policy $\pi_{\theta_{\mathrm{po}}}$, parameterized by $\theta_{\mathrm{po}}$, and a value function $V^{\pi_{\theta_{\mathrm{po}}}}_{\phi_{\mathrm{po}}}$, parameterized by $\phi_{\mathrm{po}}$, that approximates the expected cumulative reward.
    For action selection, the adaptive retransmission agent collects channel conditions and sample-level features for representation extraction.
    The actor network outputs the retransmission decision, while the critic network estimates the corresponding state value, jointly guiding policy learning and convergence.
  \subsubsection{State and Action Formulation}
    To comprehensively characterize both the channel conditions and the feature distribution of the current sample, the for i-th image $p_i$ we define the input state $s_i$ as
    \begin{equation}
      s_i = \left\{ 
      \mathrm{SNR},\;
      \mathrm{R},\;
      \mathrm{LPIPS}_{\mathrm{th}},\;
      \mathrm{LPIPS}_{\mathrm{est}},\;
      \mathrm{LPIPS}_{\mathrm{diff}},\;
      \mathbf{\tilde{x}}_{\mathrm{com}}
      \right\},
    \end{equation}
    where $\mathrm{LPIPS}_{\mathrm{diff}}$ denotes the deviation between the estimated LPIPS $\mathrm{LPIPS}_{\mathrm{est}}$ and the target threshold $\mathrm{LPIPS}_{\mathrm{th}}$, which is calculated as
    \begin{equation}
      \mathrm{LPIPS}_{\mathrm{diff}} = \mathrm{LPIPS}_{\mathrm{est}} - \mathrm{LPIPS}_{\mathrm{th}}.
    \end{equation}
    Given the differences in scale and dimensionality among these state variables, we group them according to their physical meanings and processing each group through independent linear layers for latent-space projection and dimensional alignment.
    The resulting latent representations are concatenated and then further processed to extract high-level features.
    Within the actor network, these features are used to generate the retransmission decision. For the current sample $p_i$, the action space $A_i$ is defined as:
    \begin{equation}
      \mathcal{A_\mathrm{i}} = \{0, 1\},
    \end{equation}
    where the two actions correspond to accepting the current transmission result or requesting a retransmission, respectively. Based on the actor's output, the receiver sends a NAK signal to the transmitter to trigger retransmission encoding and transmission.

  \subsubsection{Reward Design}
    For reward function design, when multiple optimization objectives, e.g. reconstruction quality and transmission overhead,
    are highly coupled and characterized by heterogeneous scales,  
    achieving effective joint optimization is challenging. Most existing studies adopt weighted reward functions to balance the impact of various optimization objectives on policy convergence.
    However, improper weight selection in such weighted-sum designs can severely hinder policy learning and convergence,
    making it difficult to obtain an effective and robust policy.
    
    To address this issue, we adopt a sparse reward design. By constructing reward signals based on the retransmission results and the retransmission decisions, 
    the proposed approach enables policy learning to directly target transmission reliability while enhancing policy robustness.
    Transmitted samples are categorized according to whether the reconstructed quality meets the target threshold and whether retransmission effectively improves quality.
    Rewards are then assigned based on the corresponding category, encouraging the agent accurately identify severely corrupted images while minimizing overall retransmission overhead.
    For the current image $p_i$, the reward associated with action $a_i$ is defined as
    \begin{equation}
    r_i =
    \left\{
    \begin{aligned}
    10.0, \;& \mathrm{LPIPS}(p,\tilde{p}) > \mathrm{LPIPS}_{\mathrm{th}}, \\
          & \mathrm{LPIPS}(p,\tilde{p}_2) \le \mathrm{LPIPS}_{\mathrm{th}},\;
            a_i = 1 \\[4pt]
    0.5, \;& \mathrm{LPIPS}(p,\tilde{p}) \le \mathrm{LPIPS}_{\mathrm{th}},\;
            a_i = 0 \\[4pt]
    -0.5, \;& \mathrm{LPIPS}(p,\tilde{p}) > \mathrm{LPIPS}_{\mathrm{th}}, \\
          & \mathrm{LPIPS}(p,\tilde{p}_2) > \mathrm{LPIPS}_{\mathrm{th}},\;
            a_i = 1 \\[4pt]
    -5.0, \;& \mathrm{LPIPS}(p,\tilde{p}) > \mathrm{LPIPS}_{\mathrm{th}},\;
            a_i = 0 \\[4pt]
    -1.0, \;& \mathrm{LPIPS}(p,\tilde{p}) \le \mathrm{LPIPS}_{\mathrm{th}},\;
            a_i = 1
    \end{aligned}
    \right..
    \end{equation}
    The reward is determined jointly by the reconstruction quality before retransmission, the reconstruction quality after retransmission, and the agent's action.
    Specifically, higher rewards are assigned when retransmission successfully improves samples that violate the LPIPS threshold, whereas penalties are imposed for redundant or ineffective retransmission,
    as well as for failing to request retransmission for severely corrupted images.

\subsubsection{Critic Network Optimization}

The critic network is employed to approximate the state-value function under the current policy $\pi_{\theta_{\mathrm{po}}}$, which is defined as
\begin{equation}
V^{\pi_{\theta_{\mathrm{po}}}}_{\phi_{\mathrm{po}}}(s_t)
=
\mathbb{E}_{\pi_{\theta_{\mathrm{po}}}}
\left[
\sum_{k=0}^{\infty}
\gamma^{k} r_{t+k}
\;\middle|\;
s_t
\right],
\end{equation}
where $\phi_{\mathrm{po}}$ denotes the parameters of the critic network.
To train the critic, the temporal-difference (TD) error is first defined as
\begin{equation}
\delta_t
=
r_t
+
\gamma (1 - d_t)\,
V_{\phi_{\mathrm{po}}}(s_{t+1})
-
V_{\phi_{\mathrm{po}}}(s_t),
\end{equation}
where $d_t \in \{0,1\}$ indicates whether the episode terminates at time step $t$.
Based on the Generalized Advantage Estimation (GAE) framework, the advantage function is computed as
\begin{equation}
\hat{A}_t
=
\sum_{l=0}^{\infty}
(\gamma \lambda)^l \,
\delta_{t+l},
\end{equation}
where $\lambda \in [0,1]$ controls the bias-variance trade-off.
Accordingly, the regression target for the critic network is given by
\begin{equation}
\hat{R}_t
=
\hat{A}_t
+
V_{\phi_{\mathrm{po}}}(s_t),
\end{equation}
which serves as an estimate of the empirical return. The critic is trained by minimizing the MSE loss between the predicted state value and the regression target
\begin{equation}
\mathcal{L}_{\mathrm{critic}}(\phi_{\mathrm{po}})
=
\mathbb{E}_{s_t \sim \pi_{\theta_{\mathrm{po}}}}
\left[
\left(
V_{\phi_{\mathrm{po}}}(s_t)
-
\hat{R}_t
\right)^2
\right].
\end{equation}
Optimizing this loss yields a low-variance, bias-controlled estimate of the state-value function.

\section{Experiment Results}\label{simulation_results}
    In this section, we evaluate the proposed S3CHARQ framework in terms of image reconstruction performance and transmission reliability.
\subsection{Experiment Settings}
    \subsubsection{Datasets}
      In the simulation experiments, we evaluate both low-resolution and high-resolution image datasets to comprehensively validate the generality of the proposed S3CHARQ framework under diverse transmission configurations.
      It is worth noting that, due to the discrepancies in LPIPS estimation accuracy between training and testing datasets, we deliberately adopt an additional dataset during reinforcement training.
      Directly training the PPO agent on the original training set makes it difficult to capture the true deviation distribution, leading to performance degradation during deployment.
      For low-resolution image experiments, we use the CIFAR-10 image dataset as the experimental data for training and testing the encoders and decoders, while the CIFAR-100 test set is employed to train the PPO algorithm.
      Both CIFAR-10 and CIFAR-100 consist of 32 $\times$ 32 RGB images, with 50,000 training samples and 10,000 test samples.
      
      For the high-resolution image experiments, to accelerate training, we follow prior work in the selection of training dataset\cite{leung2023padc}. 
      Specifically, the ImageNet validation set is used to train the encoders and decoders, as it contains 50,000 high-resolution images with diverse content.
      The DIV2K dataset, consisting of 800 training images and 100 test images, is employed to training and testing the PPO algorithm, respectively. 
    
    \begin{figure*}[!t]
      \centering
      \includegraphics[width=\linewidth,clip,trim={1.0cm 1.0cm 1.0cm 1.0cm}]{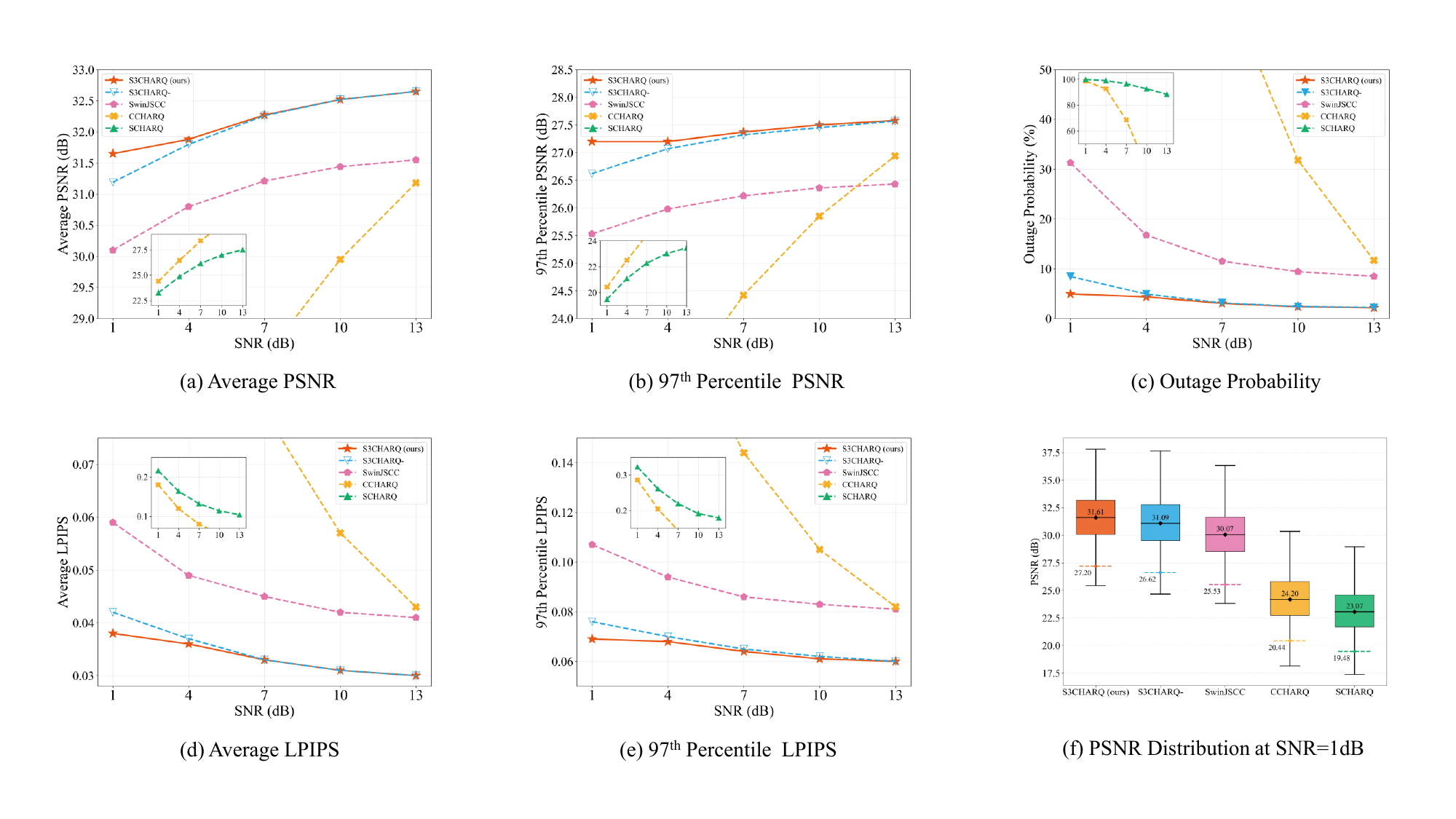}
      \caption{\small{Performance Comparison on CIFAR10 with $R=R_2=1/8$ under an AWGN channel.}}
      \label{fig:1_8_AWGN}
    \end{figure*}

    \subsubsection{Benchmarks}
    To comprehensively demonstrate the superiority of the proposed S3CHARQ algorithm, we introduce the following benchmark methods for comparative evaluation. 
    \begin{itemize}
      \item \textit{SwinJSCC}\cite{Yang2025swin}: The baseline framework of S3CHARQ, which represents an open-source JSCC approach for wireless image transmission without retransmission. 
      \item \textit{CCHARQ}\cite{cai2025ccharq}: A classic HARQ-based semantic retransmission system, which uses a single JSCC encoder for both the initial transmission and retransmission when required, and an SSIM estimator for quality prediction. To reduce transmission overhead, a DDQN network is employed to guide the progressive transmission of encoded features under a predefined SSIM threshold.
      \item \textit{SCHARQ}\cite{jiang2022sim32}: Another HARQ-based system that adopts multiple encoder-decoder pairs to independently optimize the initial transmission and retransmission. For reliability validation, SCHARQ follows conventional CRC-based checking procedure used in bit-level communications. 
      When retransmission is required, the retransmission encoder generates features with the same code length as the initial transmission. At the receiver, SCHARQ performs joint decoding using all received features.
      \item \textit{S3CHARQ$^{-}$}: A variant of the S3CHARQ framework, where the PPO-based retransmission decision module is replaced by the same rule-based strategy used in CCHARQ.   
    \end{itemize}
    \begin{table}[!t]
\centering
\caption{Neural Network Architecture of S3CHARQ}
\label{tab:s3c_harq_arch}
\renewcommand{\arraystretch}{1.15}
\setlength{\tabcolsep}{4pt}
\resizebox{\columnwidth}{!}{%
\footnotesize
\begin{tabular}{|c|c|c|}
\hline
\textbf{Network} & \textbf{Layers for Low Resolution} & \textbf{Layers for High Resolution} \\
\hline

\begin{tabular}[c]{@{}c@{}}
Backbone
\end{tabular}
&
\begin{tabular}[c]{@{}c@{}}
SwinJSCC \cite{Yang2025swin} $32\times32$
\end{tabular}
&
\begin{tabular}[c]{@{}c@{}}
SwinJSCC $256\times256$
\end{tabular}
\\
\hline

\begin{tabular}[c]{@{}c@{}}
SFA-Check \\
Encoder
\end{tabular}
&
\begin{tabular}[c]{@{}c@{}}
FC, 256 $\rightarrow$ 48 \\
5$\times$5 CNN, 256, stride=1 $\times 4$ \\
5$\times$5 CNN, 64, stride=1 \\
AF block \cite{leung2023padc} $\times 2$ \\
FC $\times 3$, 1024 $\rightarrow$ 256 $\rightarrow$ 64
\end{tabular}
&
\begin{tabular}[c]{@{}c@{}}
FC, 320 $\rightarrow$ 48 \\
5$\times$5 CNN, 256, stride=1 $\times 4$ \\
5$\times$5 CNN, 256, stride=1 \\
AF block $\times 2$ \\
FC $\times 3$, 3072 $\rightarrow$ 1024 $\rightarrow$ 256
\end{tabular}
\\
\hline

\begin{tabular}[c]{@{}c@{}}
LPIPS \\
Estimator
\end{tabular}
&
\begin{tabular}[c]{@{}c@{}}
FC, 256 $\rightarrow$ 48 \\
FC-ResBlock \cite{leung2023padc}, 3456 $\rightarrow$ 3072 \\
FC-ResBlock, 3072 $\rightarrow$ 3072 \\
FC, 3072 $\rightarrow$ 1 \\
Sigmoid
\end{tabular}
&
\begin{tabular}[c]{@{}c@{}}
FC, 320 $\rightarrow$ 48 \\
FC-ResBlock, 12672 $\rightarrow$ 3072 \\
FC-ResBlock, 3072 $\rightarrow$ 3072 \\
FC, 3072 $\rightarrow$ 1 \\
Sigmoid
\end{tabular}
\\
\hline

\begin{tabular}[c]{@{}c@{}}
Temporal \\
Entropy Model
\end{tabular}
&
\begin{tabular}[c]{@{}c@{}}
FC $\times 2$, 512 \\
Conv1D, 128 $\times 3$
\end{tabular}
&
\begin{tabular}[c]{@{}c@{}}
FC $\times 2$, 320 \\
Conv1D, 256$\rightarrow$128$\rightarrow$256
\end{tabular}
\\
\hline

\begin{tabular}[c]{@{}c@{}}
CA-JSCC \\
Decoder
\end{tabular}
&
\begin{tabular}[c]{@{}c@{}}
FC, 257 $\rightarrow$ 256 \\
SwinJSCC Decoder
\end{tabular}
&
\begin{tabular}[c]{@{}c@{}}
FC, 321 $\rightarrow$ 320 \\
SwinJSCC Decoder
\end{tabular}
\\
\hline
\end{tabular}
}
\end{table}
    \subsubsection{Implementation Details}
    To ensure a fair comparison under an identical estimation metric. For CCHARQ and SCHARQ, we replace the original SSIM estimator and CRC-based checking procedure with the same LPIPS estimator in S3CHARQ. For threshold-based retransmission schemes (CCHARQ, SCHARQ, and S3CHARQ$^{-}$), to mitigate variations in retransmission ratios caused by distribution mismatches between ground-truth and estimated LPIPS, we adopt a scaled LPIPS threshold to better align the retransmission ratios.
    Moreover, to match sample-level transmission overhead across all benchmarks, we use the same compression ratio for the initial transmission and the same code length for the retransmission stage.
    Performance is evaluated using average PSNR and 97th\footnote{$97\%$ of the total samples achieve better performance than it.}-percentile PSNR; for perceptual quality, we report average LPIPS and 97th-percentile LPIPS, along with outage probability\footnote{The proportion of samples whose decoded images still fail to meet the predefined LPIPS threshold.} to access transmission reliability.

    During training, we uniformly sample the SNR from $\{0,1,4,7,10,13\}$, while the compression ratio is also uniformly sampled from [0.1, 1.0] for low resolution datasets and from [1/48, 1/6] for high resolution datasets. The check codeword length is set to $64$ for low-resolution and $256$ for high-resolution datasets, which is substantially smaller than the overheads of the transmitted JSCC codeword.
    All benchmark models are trained 1600 epochs and 200 epochs for low-resolution and high-resolution datasets respectively, using the Adam optimizer whose learning rate is set to 0.0001. For $\gamma$ in (\ref{IBloss}), we set it 0.0001 to balance multiple optimization objectives.
    Specifically, for neural network hyperparameters, we strictly follow the configurations in \cite{Yang2025swin} and \cite{cai2025ccharq} for each benchmark. The parameters of the S3CHARQ modules are summarized in TABLE~\ref{tab:s3c_harq_arch}.
    For training the adaptive retransmission agent, the linear layer used for input-state preprocessing is set to a dimension of 64, and the agent is trained for 200 epochs using the Adam optimizer with a learning rate of 0.0001 until policy convergence. All experiments are conducted on an NVIDIA RTX4090 in Ubuntu 20.04 LTS with Intel(R) Xeon(R) Platinum 8336C CPU @ 2.30GHz under 2.1.2 PyTorch environment.

\begin{figure}[!t]
\centering
\includegraphics[width=1\linewidth,clip,trim={2.8cm 5cm 3.8cm 5cm}]{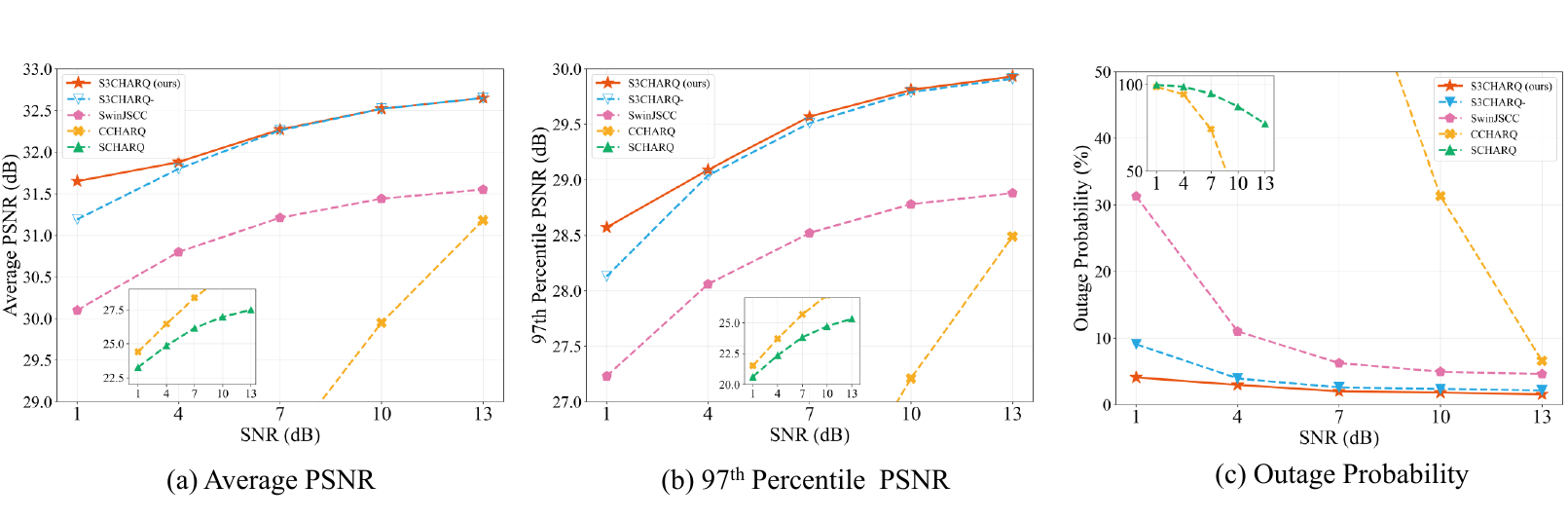}
\caption{\small{Performance Comparison on CIFAR10, with $R=R_2=1/6$ under an AWGN channel.}}
\label{fig:1_6_AWGN}
\end{figure}

\begin{figure}[!t]
  \centering
  \includegraphics[width=\linewidth,clip,trim={2.9cm 5cm 3.5cm 5cm}]{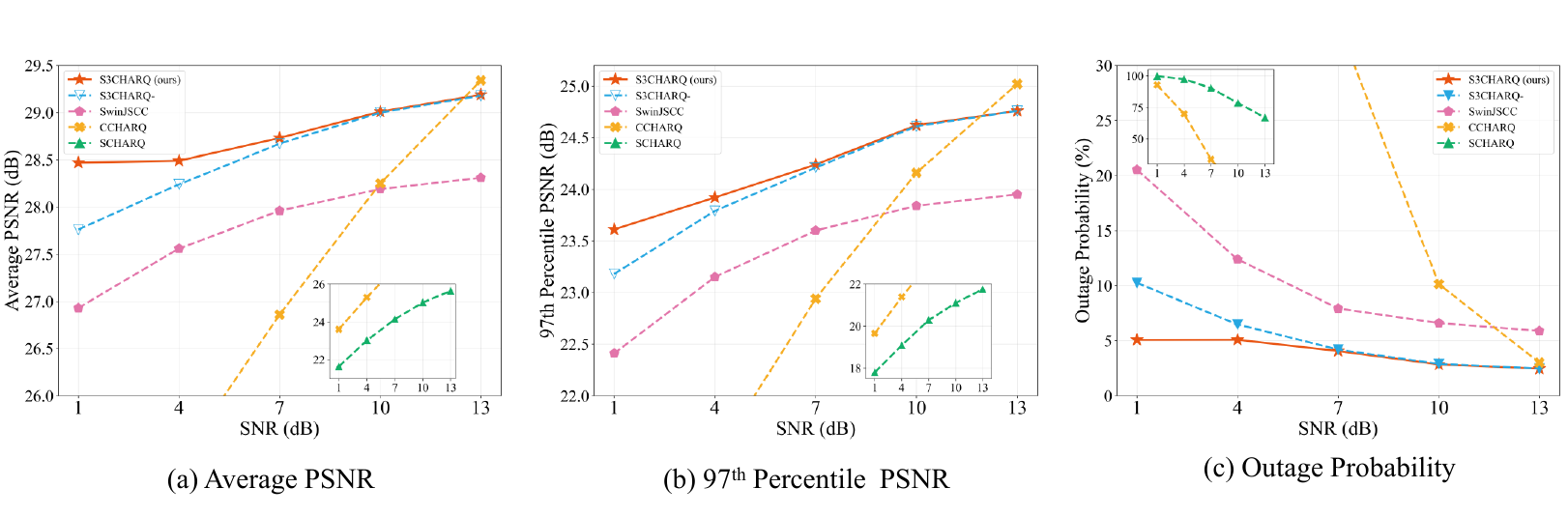}
  \caption{\small{Performance Comparison on CIFAR10, with $R=R_2=1/8$ under a Rayleigh channel.}}
  \label{fig:1_8_Rayleigh}
\end{figure}

\subsection{Validation Results}
    \subsubsection{Experiment Results on Low-Resolution Datasets}
    For the low-resolution datasets, we conduct experiments under three representative transmission settings: Setting 1: $R=R_2=1/8$ under an AWGN channel; Setting 2: $R=R_2=1/6$ under an AWGN channel, and Setting 3: $R=1/8,R_2=1/8$ under a Rayleigh fading channel.
    Regarding the retransmission threshold, we select the 90th percentile of the ground-truth LPIPS values on the test set to align the retransmission ratio across methods.
  \begin{figure*}[!t]
      \centering
      \includegraphics[width=\linewidth,clip,trim={1cm 0.6cm 1cm 1.0cm}]{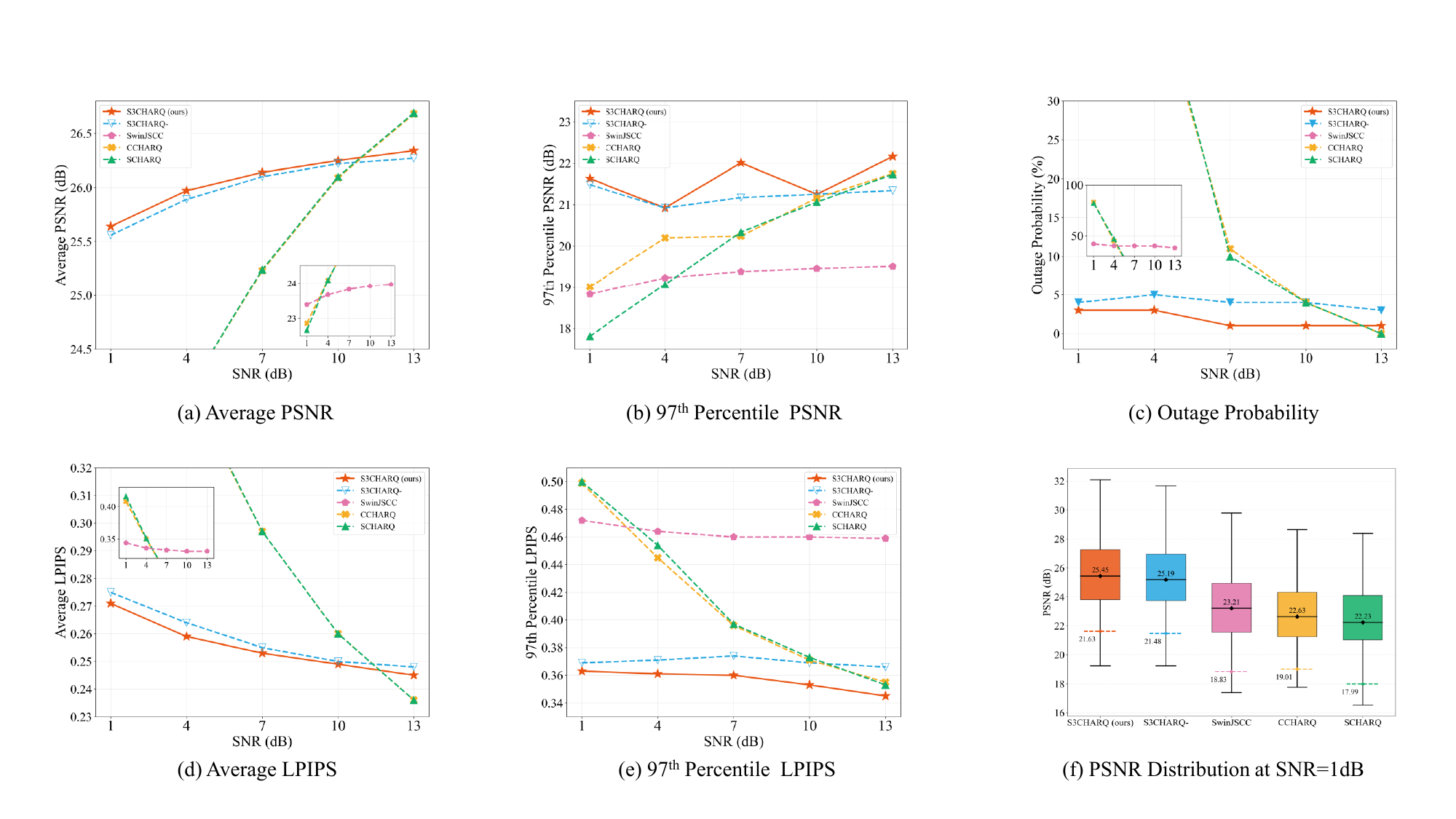}
      \caption{\small{Performance Comparison on DIV2K, where $R=R_2=1/48$ under an AWGN channel.}}
      \label{fig:1_48_AWGN}
    \end{figure*}

    \begin{figure}[!t]
      \centering
      \includegraphics[width=1\linewidth,clip,trim={2.8cm 5cm 3.6cm 5cm}]{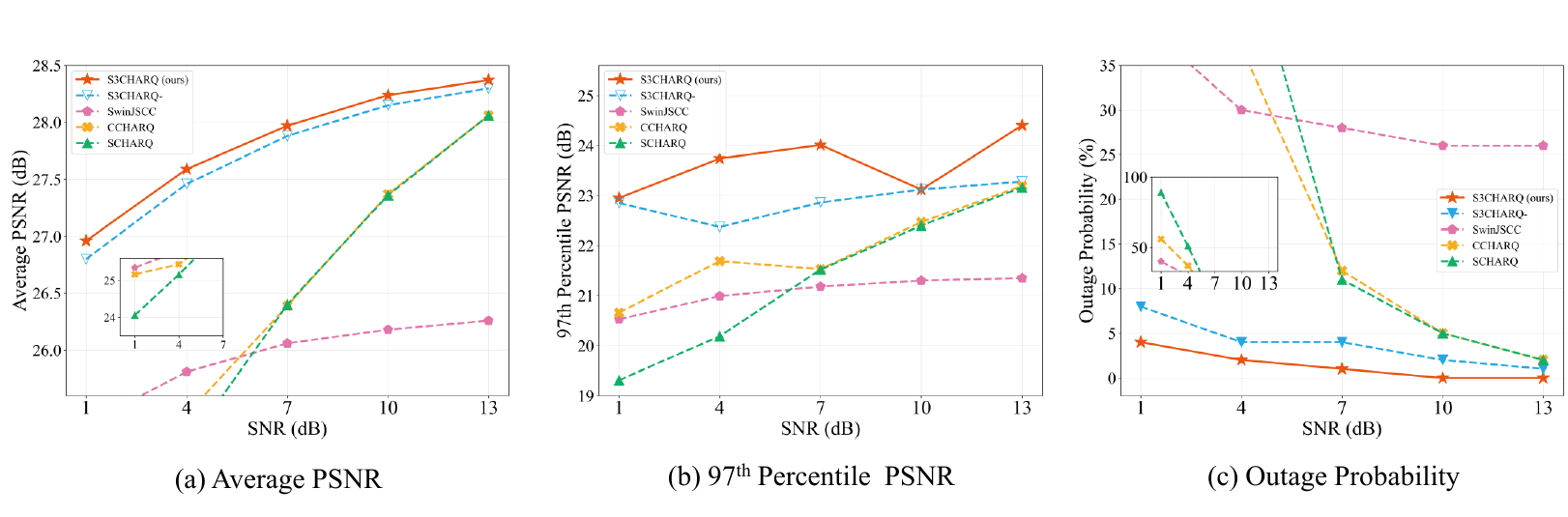}
      \caption{\small{Performance Comparison on DIV2K, where $R=R_2=1/36$ under an AWGN channel.}}
      \label{fig:1_36_AWGN}
    \end{figure}
    
    Fig.~\ref{fig:1_8_AWGN}-Fig.~\ref{fig:1_8_Rayleigh} illustrate the performance of different methods under each transmission setting.
    As shown in Fig.~\ref{fig:1_8_AWGN}(a)-(c), S3CHARQ achieves average PSNR gains of $4.12$ dB and $6.44$ dB over SCHARQ and CCHARQ, respectively.
    Compared with S3CHARQ$^{-}$, the proposed S3CHARQ attains an additional $0.46$ dB PSNR gain at $\mathrm{SNR}=1\mathrm{dB}$, demonstrating its improved transmission efficiency.
    In terms of transmission reliability, S3CHARQ consistently achieves lower outage probability across all SNR conditions.
    As shown in Fig.~\ref{fig:1_8_AWGN}(f), S3CHARQ achieves a 97th-percentile PSNR of $27.20$ dB at $\mathrm{SNR}=1\mathrm{dB}$, outperforming the variant withput RL agent by $0.58$ dB.
    Under Setting 1, S3CHARQ achieves a minimum outage probability of $3.38$\%, which is significantly lower than that of SCHARQ ($95.30$\%) and CCHARQ ($60.67$\%).
    Compared with S3CHARQ$^{-}$, S3CHARQ further reduces the outage probability by $0.85$\% overall and by $3.53\%$ at $\mathrm{SNR}=1\mathrm{dB}$.
    On 97th-percentile PSNR performance, as shown in Fig.~\ref{fig:1_8_AWGN}(f), S3CHARQ achieves $27.20$ dB at $\mathrm{SNR}=1\mathrm{dB}$, outperforming variant without RL Agent by $0.58$ dB.
    For perceptual quality, as shown in Fig.~\ref{fig:1_8_AWGN}(d)-(e), S3CHARQ achieves the best LPIPS performance aming all methods, with an average LPIPS of $0.041$ and a 97th-percentile LPIPS of $0.064$.
    These results clearly demonstrate the effectiveness of the proposed S3CHARQ framework in improving both transmission efficiency and reliability. 

    Similarly, as shown in Fig.~\ref{fig:1_6_AWGN}, S3CHARQ continues to demonstrate superior performance at lower compression ratios in terms of average PSNR, 97th-percentile PSNR, and outage probability. 
    For instance, S3CHARQ achieves an outage probability of only $2.50\%$, outperforming S3CHARQ$^{-}$ ($4.02\%$), CCHARQ ($53.23\%$), and SCHARQ ($91.17\%$). 
    As the transmission overhead constraint is relaxed, the advantage of S3CHARQ in transmission reliability becomes even more pronounced.
    These results further demonstrate the effectiveness of jointly optimizing the encoding and check coding procedures in improving both transmission efficiency and reliability.
    
    For experiments under Rayleigh fading channels, as shown in Fig.~\ref{fig:1_8_Rayleigh}, at $\mathrm{SNR}=13\mathrm{dB}$ condition, CCHARQ achieves gains of $0.15$ dB in Average PSNR and $0.26$ dB in 97th-percentile PSNR, compared with S3CHARQ. 
    Unlike the AWGN case, even at relatively high average SNR, instantaneous fading-induced fluctuations introduce severe noise interference to the transmitted features. Compared with CCHARQ, which encodes features once and transmits them progressively, S3CHARQ re-encodes features during retransmission based on feedback, 
    which may further propagate transmission errors under fading conditions, leading to slightly degraded average and tail PSNR performance. 
    Overall, under the same compression ratio and retransmission code length, S3CHARQ achieves superior PSNR and LPIPS performance in most settings compared with other benchmarks, indicating higher encoding efficiency.
    In terms of transmission reliability, S3CHARQ consistently maintains lower outage probability than competing schemes, effectively ensuring a lower bound on reconstruction quality.
    Moreover, the gains in efficiency and reliability remain consistent across both low- and high-SNR regimes, confirming that S3CHARQ enables more refined and reliable transmission encoding.

    \subsubsection{Experiment Results on High-Resolution Datasets}
    Similar to the low-resolution experiments, for high-resolution evaluation, we consider two representative compression settings, namely Setting 1: $R=R_2=1/48$ under an AWGN channel, and Setting 2: $R=R_2=1/36$ under an AWGN channel.
    Specifically, during training and validation, all original images are resized to a standard resolution of 256 $\times$ 256.
    This resizing ensures a consistent input size for the JSCC encoder across all benchmarks and improves training efficiency.
    
    As shown in Figs.~\ref{fig:1_48_AWGN} and \ref{fig:1_36_AWGN}, S3CHARQ outperforms all benchmark methods under most SNR conditions,
    demonstrating stable and superior performance across multiple metrics, including average PSNR, 97th-percentile PSNR, outage probability, as well as average and 97th-percentile LPIPS.
    It is worth noting that under Setting 1, although CCHARQ and SCHARQ achieve $0.34$ dB higher average PSNR than S3CHARQ at $\mathrm{SNR}=13\mathrm{dB}$,
    S3CHARQ still provides significant gains in tail performance, delivering improvements of $1.12$ dB and $1.59$ dB in 97th-percentile PSNR over CCHARQ and S3CHARQ, respectively.
    For perceptual quality, as shown in Fig.~\ref{fig:1_48_AWGN}(d)-(e), S3CHARQ achieves an average LPIPS of $0.255$ and a 97th-percentile LPIPS of $0.356$,
    significantly outperforming CCHARQ ($0.310$ on average, $0.413$ at the 97th-percentile) and SCHARQ ($0.312$ on average, $0.415$ at the 97th-percentile).

    For experiments conducted under Setting 2, as illustrated in Fig.~\ref{fig:1_36_AWGN}, 
    S3CHARQ maintains stable performance advantages with less strigent compression and retransmission overhead constraints. 
    Compared with other benchmarks, S3CHARQ achieves an outage probability of $1.40\%$, demonstrating its  strong capability in ensuring transmission reliability.
    These results highlight the effectiveness of optimizing information redundancy across different coding procedures to improve reliability.
    Moreover, as communication resources increase, the performance gains enabled for this optimization become even more pronounced.

    \subsubsection{Visualization Results}
    
    \begin{figure}[!t]
      \centering
      \includegraphics[width=\linewidth, clip, trim={1.2cm 0 14.3cm 0}]{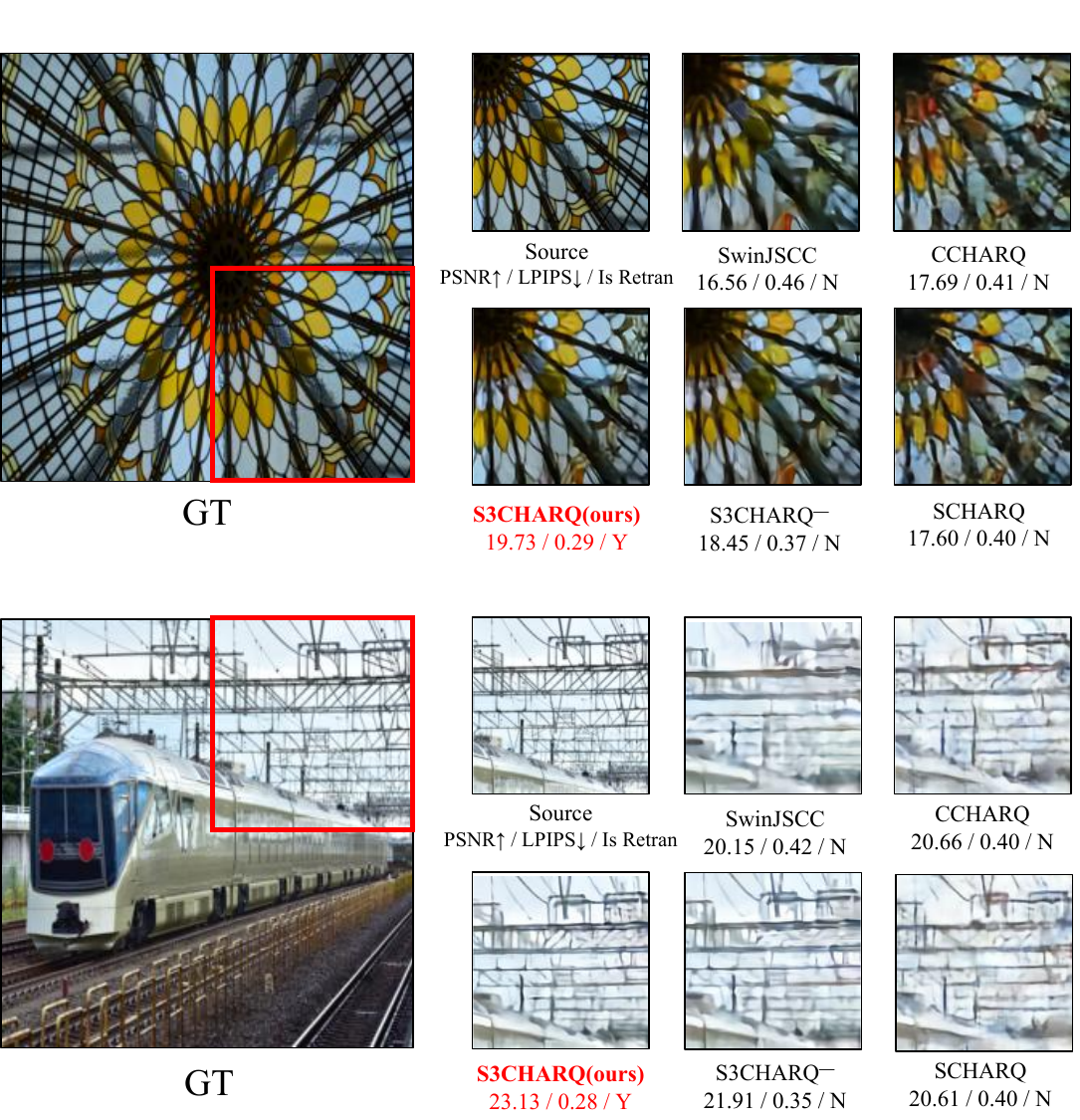}
      \caption{\small{Visualization results on DIV2K.}}
      \label{fig:Visualization}
    \end{figure}
    As shown in Fig.~\ref{fig:Visualization}, to further illustrate the effectiveness of the proposed S3CHARQ, we present a set of qualitative visualization results on the DIV2K dataset.
    The visual comparisons are conducted under an AWGN channel with $\mathrm{SNR}=4\mathrm{dB}$, a compression ratio of $R=R_2=1/48$, and a scaled retransmission threshold set to $0.339$. 

    As observed from the visual results, after the first transmission round, the reconstruction quality remains well below the preset LPIPS threshold.
    All rule-based benchmarks fail to detect this perceptual degradation due to estimation errors between the predicted and true LPIPS values.
    For example, for the first image, CCHARQ decides not to retransmit based on an estimated LPIPS of $0.331$, while SCHARQ and S3CHARQ$^{-}$ yield estimates of 0.329 and 0.289, respectively.
    In contrast, instead of relying solely on the estimated LPIPS, S3CHARQ makes retransmission decisions by jointly considering sample-level features and transmission conditions, including SNR, compression ratio, and  LPIPS estimation.
    Unlike rule-based mechanisms, it can sensitively identify severely corrupted samples under adverse channel conditions and accurately request retransmission, thereby effectively improving reconstruction quality.
    This observation not only confirms the benefit of retransmission for enhancing perceptual quality but also highlights the critical role of RL-based strategy in improving the reliability and accuracy of retransmission decisions.

  \subsection{Ablation Study on Adaptive Retransmission Agent}      
      For retransmission decision-making, the core objective is to balance additional communication overhead against reliability guarantees. 
      In rule-based methods, achieving an optimal balance through manually selected retransmission thresholds or ratios is inherently difficult. 
      To address this, we conduct an ablation study in which a brute-force search is used to determine a rule-based retransmission threshold that matches the retransmission ratio learned by the RL agent under a balanced overhead-reliability tradeoff. 
      This variant is referred to as \textit{S3CHARQ$^{-}$OFF}.
      
      \begin{figure}[!t]
            \centering
            \includegraphics[width=\linewidth,clip,trim={1cm 1cm 2.3cm 1.2cm}]{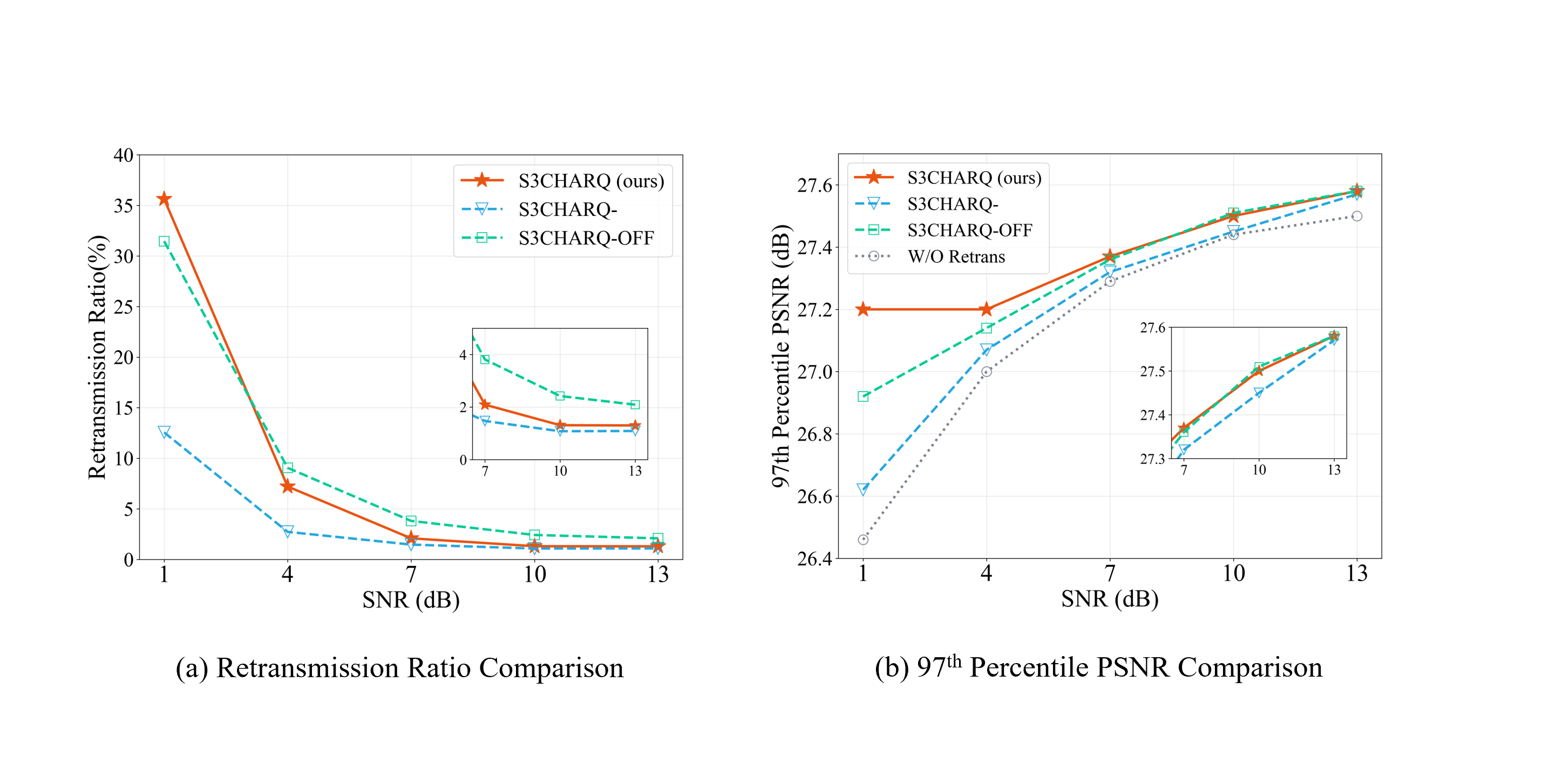}
            \caption{\small{Ablation Study on CIFAR10 under an AWGN channel.}}
            \label{fig:Ablation}
      \end{figure}

      As shown in Fig.~\ref{fig:Ablation}, the retransmission ratio of the offline rule-based scheme is largely determined by the LPIPS distribution of samples after the initial transmission, resulting in relatively uniform retransmission behavior across different SNRs. This lack of adaptivity leads to inefficient retransmission resource allocation.
      In contrast, S3CHARQ dynamically adjusts retransmission decisions according to channel conditions. 
      For instance, under low-SNR conditions, S3CHARQ tends to trigger retransmission for samples with estimated LPIPS values close to the threshold,
      mitigating misjudgments caused by estimation errors. When the SNR is higher, it adopts more conservative retransmission decisions to reduce communication overhead.
      At low SNR (1 dB), S3CHARQ achieves a $0.28$ dB gain in 97th-percentile PSNR over the offline scheme, while under high-SNR conditions, it maintains comparable reconstruction quality with consistently fewer retransmissions.
      These results highlight that by leveraging reinforcement learning, S3CHARQ enables sample-level, context-aware retransmission decisions. Such adaptive behavior effectively compensates for estimation inaccuracies and achieves a more favorable tradeoff between transmission efficiency and reliability than rule-based threshold schemes.

\section{Conclusion}\label{conclusion}
    In this paper, we proposed S3CHARQ, a semantic communication system with an adaptive retransmission mechanism for image transmission.
    To jointly optimize semantic encoding and check coding, we developed JS3C, which enables collaborative decoding by leveraging both encoded features and check codewords. 
    To address the convergence challenges arising from multiple optimization objectives, we designed a training strategy based on information bottleneck theory,
    allowing joint optimization of verification and decoding processes and effectively improving reconstruction quality.
    Furthermore, we introduced a PPO-based adaptive retransmission agent that makes sample-level retransmission decisions based on current channel conditions and estimated reconstruction quality,
    thereby mitigating the impact of LPIPS estimation errors on system reliability.
    Extensive experiments demonstrated that the proposed S3CHARQ significantly outperforms the existing schemes in terms of PSNR, LPIPS, and outage probability.
    Especially under a 1/8 retransmission ratio and an AWGN channel with $\mathrm{SNR}=1\mathrm{dB}$, S3CHARQ achieves an outage probability of only 3.38\%, 
    substantially outperforming baseline HARQ schemes.
    These results clearly demonstrate the effectiveness of S3CHARQ in enhancing both transmission efficiency and reliability.

\bibliography{main}    
\bibliographystyle{ieeetr}
\end{document}